\documentclass{article}

\usepackage[utf8]{inputenc}
\usepackage{amsmath,amssymb}
\usepackage{float}
\usepackage{graphicx}% Include figure files
\usepackage{dcolumn}% Align table columns on decimal point
\usepackage{bm}% bold math
\usepackage{hyperref}
\usepackage[colorinlistoftodos,prependcaption,textsize=small]{todonotes}

\usepackage{hyperref}
\usepackage{lineno}
\usepackage[
margin=0.75in,
right=0.75in, %%% ACTIVE ONLY FOR THE COMMENTS
]{geometry}
\usepackage{color}
\usepackage{xcolor}
\usepackage{sectsty}
\definecolor{customCol1}{rgb}{0.0,0.0,0.0}
\sectionfont{\color{customCol1}}
\subsectionfont{\color{customCol1}}
\subsubsectionfont{\color{customCol1}}
\hypersetup{colorlinks=true, linktoc=all, linkcolor=customCol1,citecolor=customCol1}
\newcommand{\ptheta}{p_{_{\theta}}}
\newcommand{\pfuse}{p_{_f}}
\newcommand{\pxi}{p_{_{\xi}}}
\newenvironment{reportAbstract}
    {\begin{center}
    {\color{customCol1} \bf Abstract} 
    \end{center}
    \quotation
    }
\usepackage{authblk}
\usepackage[sort,compress,square,numbers]{natbib}

\usepackage{multicol}

\begin{document}
\title{\color{customCol1} \bf \emph{In-silico} modeling of the micromechanics of fibrous scaffolds and stiffness sensing by cells}

\date{}% It is always \today, today,
             
\author{Dhruba Jyoti Mech}
\author{Mohd Suhail Rizvi}
\affil{\small Department of Biomedical Engineering, \\
Indian Institute of Technology Hyderabad, Kandi Sangareddy, Telangana, 502284, India.}
\maketitle

\vspace{1cm}
\begin{center}
    {\bf Corresponding author}\\
    Mohd Suhail Rizvi\\
    \vspace{3mm}
    {\bf Address}\\
    Office: C-214/C, IIT Hyderabad,\\
    Kandi, Sangareddy, Telangana 502284, India.\\
    {\bf Email: }suhailr@bme.iith.ac.in\\
    {\bf Phone: }+91 40 2301 6671 
\end{center}
\newpage
\begin{reportAbstract}
Mechanical properties of the tissue engineering scaffolds are known to play a crucial role in tissue regeneration.
% have Fibrous matrices are extensively used as scaffolds in tissue regenerative applications. 
% The understanding of the mechanical properties of the fibrous matrices and their sensing by the cells is of high importance.
Here, we have utilized discrete network and finite element models to study fibrous scaffold mechanics and its dependence on structure. We have considered two loading conditions, first, uniaxial elongation (macroscopic), and second, localized cellular forces (microscopic).
% one of the most common methods used in experiments for mechanical characterization of scaffolds, 
% and second, localized cellular forces (microscopic). % as a representation of the cellular traction.
Using these two scenarios, we have tried to establish a link between scaffold micromechanics and its macroscopic mechanical properties.
% We have compared the model results with the existing experimental measurements of the mechanical response of the electrospun scaffolds under uniaxial elongation. 
We have demonstrated that the macroscopic elastic modulus of fibrous scaffold is dependent on the sample shape, size and the degree of fiber fusion.  
Under microscopic loading conditions %, such as the localized forces applied by the cells at focal adhesions, 
the deformation of the fibrous scaffolds %with random fiber arrangement 
is anisotropic with an orientation dependent decay with distance.
% Using this we estimated the microscopic stiffness of the fibrous scaffold as experienced by the cells. 
% Further, we have explored the mechanisms of the micromechanical response of the fibrous scaffold.
% We also observe that in fibrous scaffolds the effects of the localized forces can be sensed upto a long distance transmission of the localized forces in the fibrous scaffolds.
Further, we explored the stiffness sensing under the conditions of fixed stress or fixed strain application by the cells. 
With fixed strain, we found that the stiffness sensed by a cell is proportional to scaffold’s macroscopic Young’s modulus.
% However with fixed stress, stiffness sensed by the cell depends on the cell's own elasticity. 
However, for fixed stress application, it also depends on cell’s elasticity with stiffness experienced by stiff and soft cells differing by an order of magnitude. 

The analyses demonstrate that there exists a gap between mechanical properties of the fibrous scaffolds as measured from the macroscopic testings and those sensed by the cells.
This warrants a need for extreme care during the designing and reporting of experiments involving cell-scaffold interactions.
% since stiffness sensed by the cells can differ significantly from that estimated under uniaxial, biaxial or shear loading conditions. 
The insights from this work will help in the designing of tissue engineering scaffolds for applications where mechanical stimuli are a critical factor.

\vspace{1cm}
\end{reportAbstract}
% \drawhline
% \renewcommand{\contentsname}{Outline} % ToC title
% \tableofcontents

\newpage
% \begin{multicols}{2}
\section{Introduction}
Tissue engineering has made tremendous progress in the past couple of decades towards regeneration of tissues and organs in laboratory \cite{Ringe2012,Langer2016,Rustad2010}. %Zhao2020
Tissue regeneration involves cells, scaffolds for structural support to the cells, and appropriate biochemical factors and biomechanical stimuli \cite{Howard2008,Salinas2018}. %Dey2020
The choice of cells in any particular tissue engineering application is quite limited as they are usually of the same type as the target tissue \cite{Caddeo2017Review} %Phull2016
or, in some cases, specific stem cells which differentiate to the desired cell type \cite{Howard2008}. %Ude2018
The bigger challenge is the design of a scaffold which can provide appropriate chemo-mechanical micro-environment, similar to native extracellular matrix (ECM), to the cells and also can give sufficient mechanical support in case of load bearing tissues \cite{Schiavi2018,Panadero2016}. %Choi2018

% \subsection*{Fibrous scaffolds and mechanics}
For the fibrous structure is a ubiquitous feature of the ECM \cite{Weaver2010Review}, fibrous scaffolds are  utilized in tissue engineering quite extensively \cite{Capulli2016Review,Stocco2018}. %Li2011Book,Jun2018Review,uribe2021soft,Smith2008
These scaffolds are either derived from the ECM of tissues  by the process of de-cellularization \cite{yu2016decellularized} %rajab2020decellularized
or are fabricated in lab from biological and synthetic materials using techniques such as electrospinning \cite{sill2008electrospinning,agarwal2009progress}
or three dimensional (3D) printing \cite{wang20203d}, %jammalamadaka2018recent 
with different techniques providing distinct control over scaffold architecture.
% The extent of control of the scaffold architecture is method dependent with decellularization offering almost no structural control, whereas, 3D printing offers precision, and electrospinning somewhere in between.
These fibrous scaffolds have been shown to be extremely successful in the regeneration of articular cartilage \cite{Zhao2020}, 
retina \cite{nair2021retina}, 
inter-vertebral disc \cite{choi2019ivd}, %xu2015intervertebral 
skin \cite{bhardwaj20183d_skin}, %chaudhari2016skin
and several other tissue types \cite{vasiliadis2020tendon,devillard2021vascular,ghassemi2018bone}. %pankajakshan2010vascular

% \subsection*{Mechanical characterization of scaffolds}
% Therefore, the mechanical characterization of tissue engineering scaffolds is one of the most important steps in any tissue engineering exercise. 
% For these applications an understanding of the scaffold mechanics is very crucial, more so in case of the load bearing tissues where mechanical microenvironment is a critical factor \cite{chen20153d,hao2017scaffold}.
Owing to the importance of the mechanical forces in tissue engineering \cite{liebschner2005mechanical} %Oddou2005
the characterization of the mechanical properties of the fibrous scaffolds is a common feature in almost all tissue engineering exercises. 
It is usually performed by measuring its response against several loading conditions such as uniaxial elongation, 
biaxial elongation
shear deformation
or dynamic deformation for viscoelastic properties \cite{buffinton2015uniaxialElongation,sacks2000biaxial,ghezelbash2021shear}.
These measurements provide an insight into the overall macroscopic mechanical properties of the fibrous scaffold.
This, however, does not ensure that the cells seeded onto the scaffold also experience the same mechanical response against microscopic cellular forces/deformations.
It is therefore important to reliably predict the mechanical environment cells are going to encounter when seeded in a particular fibrous scaffold. 

% \subsection*{Macroscopic modeling approaches: discrete vs continuum}
In addition to the experimental methods, mathematical and computational approaches are extensively utilized to study the mechanical response of the fibrous scaffolds.
One of the most prominent method for the mathematical modeling of the mechanical response of any material is the constitutive modeling under continuum assumption \cite{ogden1997non}.
This approach has also been utilized for the fibrous materials to formulate structure based constitutive models of ECM \cite{HolzapfelGasserOgden2000,FedericoGasser2010} %Gasser2006 
and fibrous scaffolds \cite{Shenoy2014b}. 
These continuum based approaches, however, by definition, consider the microscopic length scale of the scaffold to be much smaller than the length scale of interest. 
This assumption is valid for the macroscopic uniaxial/biaxial or shear deformations of the scaffold but for the study of the cell-scaffold interactions continuum assumption is questionable \cite{Jones2015PNAS}.
This has led to the formulation of several statistical description based models of fibrous scaffolds \cite{Rizvi2012ActaBiomat,Rizvi2014JMBBM,Rizvi2016BMM} and discrete fiber network models \cite{Razavi2020,Eichinger2021,Tyznik2019,MacKintosh2015,Rohanifar2020,JanmeyShenoy2019}. 
These works have looked at different aspects of the nonlinear mechanics of the fibrous networks including length scale dependent mechanics \cite{Tyznik2019}, effect of individual fiber mechanics \cite{Rohanifar2020}, anomalous Poisson effects \cite{JanmeyShenoy2019}.

Further, the effect of fusion of the fibers is not considered in the continuum constitutive models  \cite{FedericoGasser2010}, whereas, in the discrete fiber models the all the fiber-fiber contacts are usually considered to be perfectly fused \cite{Razavi2020}.
In some methods of fibrous scaffold fabrication, such as electrospinning, the fiber fusion can be controlled by the operating conditions \cite{raghavan2011control}.
Therefore, the study of the mechanical behavior of the mechanics of fibrous scaffolds need to take into account the degree of fiber fusion. 

% \subsection*{Micromechanics}
Despite a lot of work on the macroscopic rheology and constitutive modeling under uniaxial and shear loading conditions, the fibrous scaffold mechanics at the cellular length scale (or microscopic scale) has remained less explored. 
Recently, there has been a lot of activity in this direction to understand the micromechanics of fibrous networks against localized cellular forces
\citep{Jones2015PNAS,Grimmer2018,Proestaki2019,Goren2020BiophysJ,Mann2019,Tyznik2019,Eichinger2021}. % Burkel2018PRE, Burkel2017,Rohanifar2020
Experimental and modeling studies have shown non-affine \cite{Burkel2018PRE} and anisotropic \cite{Goren2020BiophysJ} deformations of fibrous scaffolds under cellular forces.
Computational modeling has also shown force chain formations \cite{Mann2019} and long range force transmission in the fiber networks under localized forces \cite{Hall2016PNAS,Broedersz2018,Shenoy2014}.
% Furthermore, some of these works on the micromechanics of fibrous networks have mainly focused on the network deformation \cite{Shenoy2014} and transmission of forces \cite{Shenoy2014b}.
However, the nature of the sensing of the mechanical properties of the fibrous scaffolds, such as stiffness, by the cells has not been looked into in any considerable detail.
One of the possible reason for this is the lack of mechanistic understanding of the mechanosensing by the cells where it is not yet clear if cells apply stresses or strains to the substrate \cite{DeSafran2008,Panzetta2019PNAS}.

In this work, we have looked at the micromechanics of fibrous scaffolds and its relationship with its macroscopic elastic properties and structure. 
We formulate a discrete fiber network computational model (DFM) using an approach motivated by the electrospinning process \cite{Electrospinning2019Review}
and compare its results against experimental measurements and also continuum constitutive model of fibrous materials.
We also looked at the stiffness sensing of the fibrous scaffolds by the cells and its dependence on the cell's mechanical properties and the mechanosensing mechanism.  
%\emph{in situ} conditions of the cell in scaffold \emph{in silico} [REF].
\section{Model Description}
\subsection{Discrete fiber model}
Application of mechanical forces, either in mechanical testing setup or by cells via focal adhesions, results in the deformation of the fibrous scaffold. 
This scaffold deformation results in the deformation of the fibers in the form of their bending, straightening, and buckling.
Further, depending on the fabrication conditions the fibers in the scaffold can also fuse with each other \cite{raghavan2011control}.
These inter-fiber fusion points can transmit the mechanical load from one fiber to another when the scaffold undergoes deformation.
In order to study the mechanical behavior of fibrous scaffold as realistically as possible, therefore, it is essential to have its detailed description in terms of the fiber geometry and fiber-fiber fusion locations. Towards this goal, we describe the scaffold in terms of its constituent fibers as detailed next.
\subsubsection{\emph{In-silico} fibrous scaffold generation}
\begin{figure}[H]
    \centering
    \includegraphics[width=0.8\textwidth]{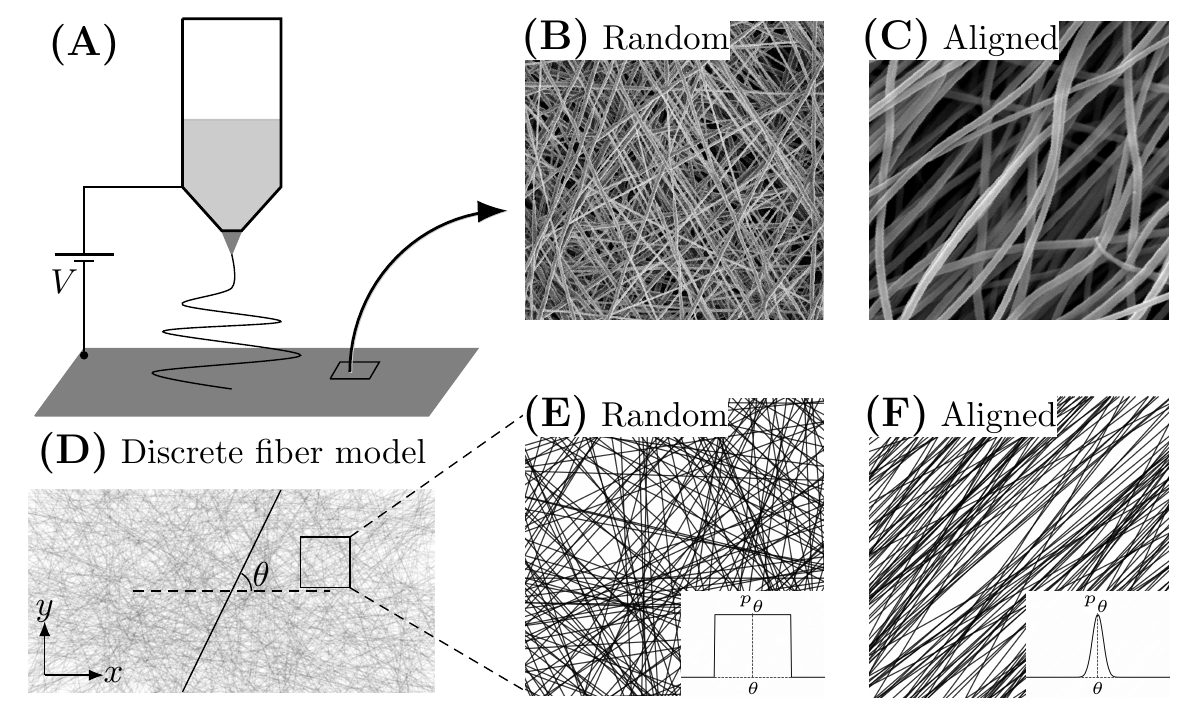}
    \caption{{\bf Electrospinning and discrete fiber model.} (A) Schematic of electrospinning setup and sample electron micrographs of (B) random (composed of Polycaprolactone/Gelatin fibers) and (C) aligned (composed of polystyrene fibers) electrospun fibrous matrices. (D) Generation of discrete fiber model by deposition of individual fibers in the form of straight line segments. (E) Random and (F) aligned discrete fiber matrices generated using two different probability distributions $\ptheta$ (shown in respective insets).}
    \label{fgr:schematic}
\end{figure}
The discrete fiber model is motivated from the electrospinning method of fibrous scaffold fabrication. 
In electrospinning, a liquid jet is generated from an electrified droplet (Fig. \ref{fgr:schematic}A). 
With the help of the electric field the jet is stretched and elongated to generate fibers which are deposited at the collector surface  \citep{Electrospinning2019Review,Luraghi2021Review}. 
Owing to the bending instabilities in the extruded jet the deposited fibers at the collector are randomly arranged (see Fig. \ref{fgr:schematic}B for reference). 
However, the arrangement of the fibers in electrospinning can be modulated to an aligned arrangement (see Fig. \ref{fgr:schematic}C), upto a point though, by deploying a rotating collector \citep{Electrospinning2019Review}. 

The electrospun fibrous matrices usually have a very small thickness \cite{ryu2020thickness}. Therefore, we develop a two dimensional model of the fibrous scaffold by following the same approach and `deposit' virtual fibers, in the form of straight line segments (for scaffold with straight fibers)  onto a large region of dimensions $L_{mat} \times W_{mat}$ (Fig. \ref{fgr:schematic}D). 
These fibers are added to the virtual fibrous mat one-by-one until the total length of all the fibers does not reach a pre-specified threshold. 
This pre-specified threshold value of the fiber length corresponds to the duration of the electrospinning process or the amount of material used in the process.
We define the orientation of a fiber as the angle $\theta$ made by the line joining its two end from a pre-specified direction, as shown in Fig. \ref{fgr:schematic}D. 
Further, for a fibrous scaffold with curved fibers we characterize the geometry of a fiber in terms of its normalized length $\xi=l/R$, the ratio of its contour length and the shortest distance between its two ends.
This means for a scaffold with perfectly straight fibers, we have $\xi=1$ for all the fibers. 
For each fiber the orientation angle and length are taken from respective probability density functions (PDFs), $\ptheta (\theta)$ and $\pxi(\xi)$, which, for the results shown in this paper, are taken to be Gaussian distributions 
\begin{equation}
    \ptheta (\theta) = C_{\theta} \exp{ \left(-\frac{\left(\theta - \theta_0 \right)^2}{2\sigma_{\theta}^2} \right)} \text{ with }-\frac{\pi}{2} \le \theta \le \frac{\pi}{2},
\end{equation}
and 
\begin{equation}
    \pxi (\xi) = C_{\xi} \exp{ \left(-\frac{\left(\xi - \xi_0 \right)^2}{2\sigma_{\xi}^2} \right)} \text{ with } \xi \ge 1
\end{equation}
for fibers with no residual stresses. Here, $\theta_0$ and $\xi_0$ are the mean fiber orientation angle and mean normalized fiber length, respectively, $\sigma_{\theta}$ and $\sigma_{\xi}$ are the standard deviations of the respective PDFs, and $C_i$'s are the normalization coefficients to ensure 
\begin{equation}
    \int \limits_{-\pi/2}^{\pi/2} \ptheta (\theta) d \theta = 1 \text{ and }
    \int \limits_{1}^{\infty} \pxi (\xi) d \xi = 1.
\end{equation}
Larger and smaller values of $\sigma_{\theta}$ result in random (Fig. \ref{fgr:schematic}E) and aligned (Fig. \ref{fgr:schematic}F) fiber systems, respectively.
% Even though $\sigma_{\theta}$ can describe the degree of the alignment in the fibrous matrix, it can take arbitrarily large values. Therefore, it is useful to define a degree of fiber alignment in the matrix which can take finite values. Following the definition of orientation order parameter in nematic liquids \cite{andrienko2018introduction}, we define degree of fiber alignment as 
% \begin{equation}
% %%% https://physics.stackexchange.com/questions/65358/2-d-orientational-order-parameter
%     \mathcal{A} = %\langle 2 \cos^2 \left( \theta -\theta_0\right) - 1\rangle = %C \int \limits_{-\pi/2}^{\pi/2} \left(2 \cos^2 \theta - 1 \right) e^{-\frac{\left(\theta - \theta_0 \right)^2}{\sigma_{\theta}^2}} d \theta
%     2 C_{\theta} \int \limits_{-\pi/2}^{\pi/2} \cos^2 \left( \theta -\theta_0\right) \ptheta (\theta) d \theta - 1,
% \end{equation}
% which takes values $0$ and $1$ for fibrous matrices with random and aligned arrangement of fibers. 
Further, the fusion among the fibers is described by a probability value, $\pfuse$. 
$\pfuse$ fraction of all the fiber-fiber intersections are randomly selected in the model and the two participating fibers are considered to be fused with each other at those intersections. 
Therefore, $\pfuse=0$ and $\pfuse=1$ stand for the fiber systems with no inter-fiber fusion and fully fused fibers, respectively.  
Similar to the electrospun fibrous mat, this process generates a 2D virtual fibrous mat of dimensions $L_{mat} \times W_{mat}$. 
However, the density of the fibers in this mat is not uniform with central region of the mat having higher number of fibers relative to the mat edges \cite{ryu2020thickness}. 
Therefore, we extract the samples of required dimensions $L \times W$ (where $L, W \ll L_{mat}, W_{mat}$) from the central region of the mat for further analysis.
This discrete fiber model of the fibrous scaffold is, therefore, described in terms of the coordinates of the fiber fusion points and the geometry of the fibers connecting these points to each other and to the sample boundary (Fig. \ref{fgr:fig2_schematic_fiberLenDist}A). 
\subsubsection{Microscopic length scale of the scaffold} \label{sec:micro_length}
\begin{figure}[H]
    \centering
    \includegraphics[height=0.35\textwidth]{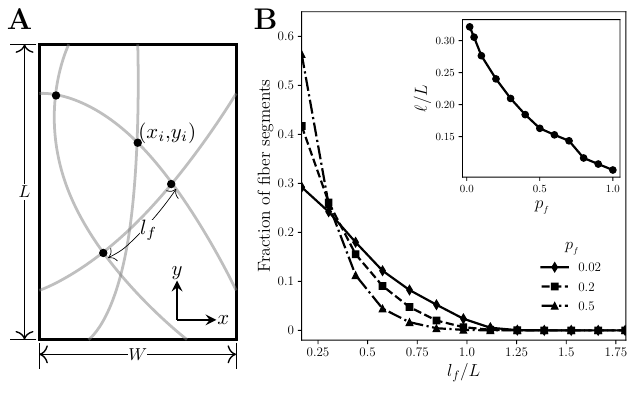}
    \caption{{\bf Microscopic length scale in fibrous scaffold}. (A) Schematic showing fiber arrangement and fiber fusion points (marked with black dots). $l_f$ is defined as the length of the fiber segment between two consecutive fusion points. (B) Distributions of $l_f$, for different $\pfuse$ for random fiber arrangement. Inset shows the dependence of $\ell = \langle l_f \rangle$, the characteristic microscopic length scale in a random fibrous scaffold, on fiber fusion probability $\pfuse$. %Here $L$ is the side length of the fibrous scaffold sample.
    }
    \label{fgr:fig2_schematic_fiberLenDist}
\end{figure}

Since in this work we are also comparing the mechanical responses of the fibrous scaffold as obtained from the discrete fiber model and continuum finite element models, it is important to define a microscopic length scale (akin to bond length or inter-particle distance in solids and mean free path in fluids \cite{gurtin1982introduction}) which can help us ascertain the validity of the continuum assumption based models for fibrous scaffolds.
For this purpose we look at the distances $l_f$ between adjacent fusion points on all the fibers. 
Fig. \ref{fgr:fig2_schematic_fiberLenDist}B shows the distribution of $l_f$ for different degrees of fiber fusion.
We utilize $\ell = \langle l_f \rangle$, the mean value of $l_f$, as the microscopic length scale of fibrous scaffold which is smaller  for higher degree of fiber fusion (inset in Fig. \ref{fgr:fig2_schematic_fiberLenDist}B). 

With this description of the fibrous scaffold geometry we will now describe the modeling of its mechanical response against macroscopic and microscopic loading conditions. 
\subsubsection{Mechanical response of a single fiber} \label{sec:eff_fiber}
Deformation of the fibrous scaffold results in bending and stretching of the fibers. 
The combined effect of the bending (or straightening) and stretching of the fibers is usually written, in a phenomenological manner, as an exponential function of the fiber strain \cite{courtney2006design}. 
In contrast to the phenomenological approach, in the following, we are going to derive a constitutive law for the effective fiber while taking into account its geometrical properties. 

For this, we are first going to look at the fiber response under stretching and bending separately. 
The application of mechanical tension on a straight fiber results in its stretching. We model fiber stretching by neo-Hookean response which is described in terms of the energy 
\begin{equation}
    u_e (\xi,\lambda_e) = \frac{1}{2} k_e \left( \lambda_e^2 + \frac{2}{\lambda_e} -3 \right)
\end{equation}
where $\lambda_e$ is the stretch ratio of the fiber and $k_e=EA$ is the product of the Young's modulus $E$ of the fiber material and fiber cross-sectional area $A$. 
This gives the force required for the stretching of a straight fiber as 
\begin{equation} \label{eq:fe_lambda_e}
    f_e (\lambda_e) = k_e\left( \lambda_e - \frac{1}{\lambda_e^2}\right).
\end{equation}

On the other hand the deformation of any curved fiber or compression of a straight fiber results in the change of the fiber curvature.
Since, in our description the fiber geometry is described in terms of a single parameter $\xi$, the normalized fiber length, we write its average curvature (in undeformed configuration) in terms of $\xi$ as (see Appendix A for details)
\begin{equation}
    \langle \kappa_0(\xi) \rangle \approx 2 \pi \sqrt{1 - \sqrt{\frac{1}{\xi}}}.
\end{equation}
Similarly, for a deformed fiber the curvature is
\begin{equation}
    \langle \kappa(\xi,\lambda_b) \rangle \approx 2 \pi \sqrt{1 - \sqrt{\frac{\lambda_b}{\xi}}}
\end{equation}
where $\lambda_b$ is the stretch ratio of the line joining the two ends of the fiber. 
This gives the bending energy of the fiber in the deformed configuration to be
\begin{equation}
    u_b (\xi, \lambda_b) = \frac{k_b}{2} \left(\langle \kappa(\xi,\lambda_b) \rangle - \langle \kappa_0(\xi) \rangle \right)^2.
\end{equation}
This gives the force required for the fiber bending (or straightening) to be 
\begin{equation} \label{eq:fb_lambda_b}
    f_b (\xi,\lambda_b) \approx -\frac{k_b}{\xi} \frac{1}{\sqrt{\lambda_b \xi}} \left( 1 - \frac{\sqrt{\sqrt{\xi}-1}}{\sqrt{\sqrt{\xi}-\sqrt{\lambda_b}}}\right).
\end{equation}
This bending force also results in a requirement of a critical load (buckling) for a straight fiber under compression. 
It has to be noted that in this formulation both of these forces $f_e$ and $f_b$ act along the line joining the two ends of the fiber. 
In general, application of mechanical force results in fiber bending (or straightening) and stretching both. 
Both of these deformations can be combined by considering a series arrangement of two springs (one corresponding to the bending deformation and other for stretching) to obtain the relation \cite{storm2005}
\begin{equation}
    \lambda = \lambda_e \lambda_b\left( f \lambda_e (f)\right)
\end{equation}
where $\lambda$ and $f$ are the stretch ratio of the fiber and the total force required for its deformation, respectively.
Here $\lambda_e(f)$ and $\lambda_b(f)$, the constitutive relations of the fiber stretching and bending, are given by equations \eqref{eq:fe_lambda_e} and \eqref{eq:fb_lambda_b}, respectively. 
Using this we can define the effective fiber response as the force required for the deformation of and average fiber connecting any two points in the scaffold as 
\begin{equation}
    \langle f (\lambda) \rangle = \int \limits_{1}^{\infty} f (\xi, \lambda) p_{\xi} (\xi) d \xi. \label{eq:avg_fiber_force}
\end{equation}
This expression relates the average force and stretch $\lambda$ in the effective fiber (see supplementary Fig. S1 for plots with different fiber geometries). % deformation in the scaffold $\lambda$ to the resistance to the deformation due to its constituent fibers . 
% Fig. S1 in supplementary information shows the force-stretch of the effective fiber for different geometrical properties of the fiber. 
We now can use this effective fiber response to estimate the mechanical behavior of the fibrous scaffold.% as the collective response of all the constituent fibers. 
\subsubsection{Mechanical response of the fibrous scaffold}
In the DFM the scaffold structure is described by the coordinates of the fiber fusion points $\mathbf{x}_i=$ ($x_i$, $y_i$) (as shown in Fig. \ref{fgr:fig2_schematic_fiberLenDist}A) and $\xi$ the normalized length of the fiber segments connecting the fusion points to each other and the sample boundary. 
We define the response of the fibrous scaffold against applied mechanical loads as the collective response of all the fibers. 
To estimate the mechanical resistance of the scaffold as a whole we calculate the forces on each fusion point in the deformed configuration by combining the forces arising due to each deformed fiber (equation \eqref{eq:avg_fiber_force}) at the fusion point. 
For $i^{\text{th}}$ fusion point, the force is, therefore, given by
\begin{equation}
    \mathbf{f}_i = \sum \limits_{\alpha \in \mathcal{N}_i} \mathbf{f}_{\alpha}
\end{equation}
where $\mathbf{f}_{\alpha}$ is the tension in the $\alpha^{\text{th}}$ fiber segment and $\mathcal{N}_i$ denotes the set of all fiber segments connected to the $i^{\text{th}}$ fusion point. 
In the numerical simulation of the DFM we calculate the forces on each fiber fusion point. 
Assuming overdamped dynamics of the system we displace each fusion point by an amount proportional to the total force acting on it until forces on all fusion points are not below a pre-specified tolerance value. 
\subsection{Continuum finite element model of fibrous materials}
There are several constitutive models of the mechanical behavior of fibrous materials and composites which are based on the continuum assumption that the microscopic length scale is negligible as compared to the experimental length scale \cite{HolzapfelGasserOgden2000,FedericoGasser2010}.
These models are widely utilized for the study of the biological and non-biological applications \cite{HolzapfelGasserOgden2000,Gasser2006,FedericoGasser2010}.
In order to compare the properties of the discrete fiber model (DFM) we also studied the mechanics of fibrous scaffolds using a continuum constitutive model.
It needs to be pointed out here that the continuum assumption may not exactly hold for the cell-scaffold interactions since the length scale $\ell$ can be comparable to the cell size. 
Still it will give us insights into the limitations of the continuum models for the study of cell-scaffold interactions.

The model developed by Holzapfel, Gasser and Ogden \cite{HolzapfelGasserOgden2000} to study of the mechanics of fibrous biomaterials is very well known and has also been incorporated in finite element structural analysis softwares, such as ABAQUS. 
We also used this continuum model analyze the scaffold mechanics to compare with the discrete fiber model.
It is worthwhile at this point to give a brief description of the continuum hyperelastic constitutive model of fibrous biomaterials (see \cite{gasser2006hyperelastic} for more details) which is described by the energy density function in terms of the strain invariants as
\begin{equation}
    U = C_{10} \left( \bar{I} - 3\right) + \frac{k_1}{2k_2} \left( \mathrm{e}^{k_2 E^2} -1\right)
\end{equation}
where $C_{10}$, $k_1$ and $k_2$ are the material parameters and 
\begin{equation}
    E = \kappa \left( \bar{I}_1 - 3 \right) + \left( 1 - 3 \kappa \right) \left( \bar{I}_4 - 1 \right)
\end{equation}
with $\kappa$ describing the nature of fiber arrangements with $\kappa=0$ and $\kappa=1/3$ standing for perfectly aligned and random fiber orientations, respectively. 
Here $\bar{I}_1$ and $\bar{I}4$ are the first and fourth invariants of right Cauchy-Green tensor and the fiber orientation tensor $\mathbf{A} = \mathbf{a} \otimes \mathbf{a}$ where vector $\mathbf{a}$ represents the fiber orientation.
Please note that in this model one can incorporate multiple fiber families but for our purposes one fiber family suffices. 
Following standard analytical techniques one can obtain the expression for the stress tensor from the energy density function. 
It has to be pointed out that this particular model does not take into account any inter-fiber interactions and, therefore, the effect of fiber fusion can not be studied with this model.
In this work we have used ABAQUS/Standard finite element package with CPS3 and CPS4R type elements to study the mechanical response of two dimensional fibrous scaffold against uniaxial elongation and localized forces. 
Similar to DFM, in the finite element (FE) model too, we kept only two loading boundaries fixed under uniaxial elongation loading and all found sample boundaries were fixed for the simulation of scaffold with localized point force.  

% \textcolor{red}{The sample is modeled in ABAQUS/Standard as 2D deformable geometry with dimension 1x1 with no thickness. 
% In this simulation, all the units were assumed to be metric (kg, mm, N, sec). In the material section, anisotropic hyperelastic material was assigned with suitable material parameters like fiber modulus, fiber dispersion, etc. 
% 
% The elements used were CPS4R and CPS3 for quad and triangular mesh respectively. The solver ignores the compressibility of the material when the current elements are used in combination with plane stress formulation. Geometric non-linearity (Nlgeom) was toggled on to account for material non-linearity. Fixed boundary conditions were defined in the initial step and the load was introduced on the sample mat in the next step. Procedure type used in all the steps are Static, General.}
\subsection{Parameters values}
In this paper we have presented the results from discrete fiber model of fibrous scaffold with random and aligned fiber arrangements. 
For random fibers we have set $\xi_0= 1.18$, $\sigma_{\xi} = 0.2$ and  $\ptheta(\theta)=1/\pi$, that is uniform distribution of the fiber orientations.
We obtained these parameter values by fitting the experimental data to the discrete fiber model (as shown in Fig. \ref{fgr:fig3_uniaxialElongation}B).
For aligned fibers, these parameters are set as $\xi_0= 1.01$, $\sigma_{\xi} = 0.01$ and $\sigma_{\theta}= {\pi}/20$, that is very narrow distribution. 
The value of $\theta_0$ for aligned fibers is mentioned in respective figure captions.

For the constitutive model of the bulk fiber material we have considered fibers to be composed of electrospun polystyrene with elastic modulus $E=3.2$GPa. 
These results, however, can be translated to fibrous scaffolds of other materials by appropriate rescaling of the Young's modulus as long as the statistical description of the scaffold structure remains same.  
In tissue engineering application the fiber diameters are usually in the range of tens of nanometers to microns (for example \cite{chen2007role}). 
Here we have taken the average fiber diameter to be $a=100$nm which sets $k_e=EA=2.5\times 10^{-5}$N.
In calculations of the scaffold's Young's modulus we have taken the thickness of the sample to be $t=0.2$mm for $L_m = 6 \times 10^6$cm for a scaffold sample dimensions $L\times W = 1$cm $\times 1$cm. 
For other values of $L_m$ the thickness is assumed to change linearly with $L_m$. 

Even though the discrete fiber model is described here in two-dimensions the fibrous scaffolds are three dimensional structures. 
Therefore, the identification of the fiber fusion points by intersection of line segments in the virtual electrospinning is bound to overestimate the number of fiber-fiber contacts.
In order to compensate for this overestimation, we do not consider very high values of $p_f$ and have taken the value of fiber fusion density to be in the range $p_f=0 - 0.2$ with its exact values mentioned in respective figure captions. 

For the continuum finite element model of the scaffold with random fiber arrangement we have set the parameters to be 
$C_{10} = 0.8$MPa, $k_1 = 1400$MPa, $k_2 = 1000$ and $\kappa = 1/3$. 
% Any deviation from these parameter values for any specific case is mentioned in  respective figure caption.
\section{Results}
\subsection{\emph{In-silico} mechanical testing}
\subsubsection{Uniaxial elongation}
\begin{figure}[H]
    \centering
    \includegraphics[width=0.95\textwidth]{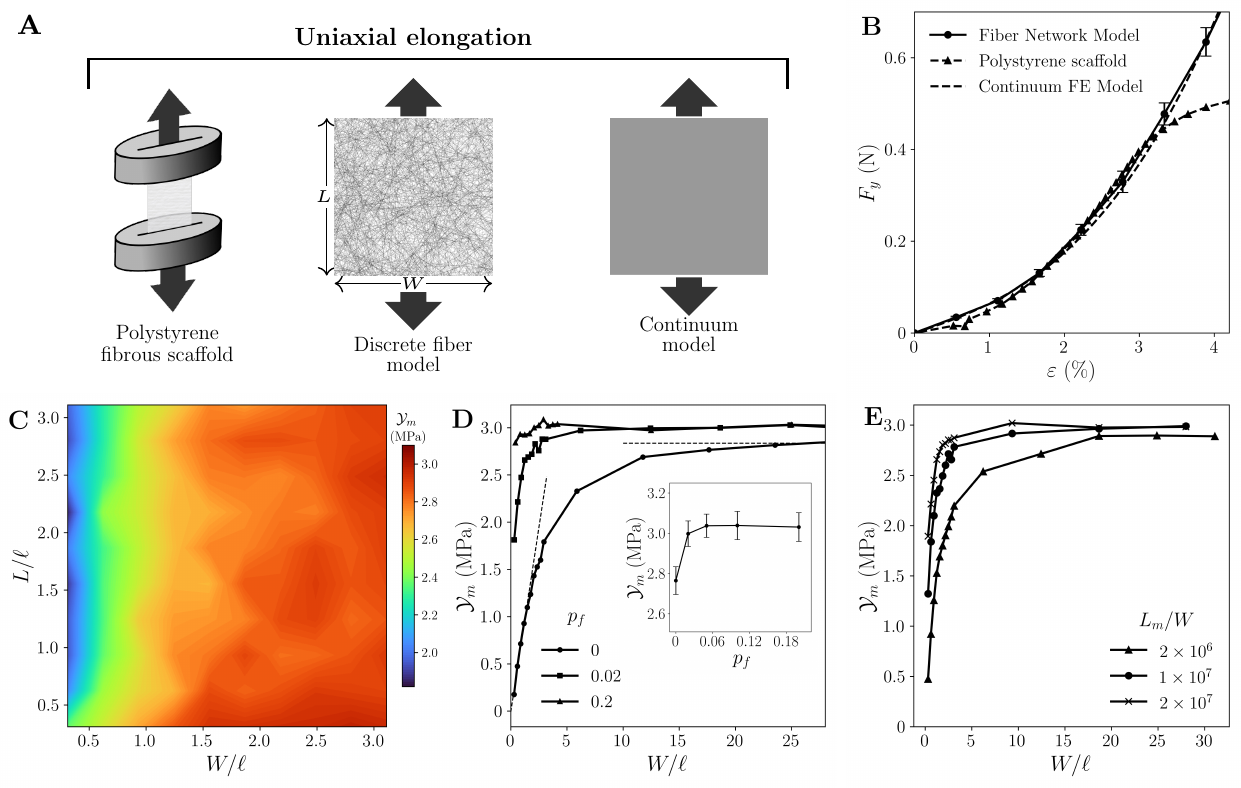}
    \caption{{\bf Uniaxial elongation of the fibrous scaffold.} (A) Schematic showing polystyrene scaffold in experiment, discrete fiber model and continuum finite element model. (B) Load-strain curves for scaffolds with random fiber arrangements as measured from mechanical testing of polystyrene scaffold (adapted from \cite{Rizvi2012ActaBiomat}), DFM ($p_f=0.02$) and FE models. Dependence of Young's modulus of fibrous scaffold on (C) sample dimensions with $L_m = 2 \times 10^7$/unit area of sample, (D) degree of fiber fusion $p_f$, and (E) $L_m$, total fibers in the sample. The dashed lines in (D) show the fit of analytical expressions for the Young's modulus, as obtained from equations \eqref{eq:Fy_eW2_smallW} and \eqref{eq:Fy_eW_largeW}. The inset in (D) shows the dependence of Young's modulus on fiber fusion density $p_f$ with $W/\ell\approx20$. The error bars in (C-E) are not shown for clarity.}
    \label{fgr:fig3_uniaxialElongation}
\end{figure}
In experiments, the mechanical behavior of the fibrous scaffolds is usually assessed by measuring the resistance of the scaffold against uniaxial elongation \cite{prasadh2018unraveling,griffin2016biomechanical}. 
Therefore, we also first analyzed the mechanical response of the fibrous scaffolds under uniaxial elongation (Fig. \ref{fgr:fig3_uniaxialElongation}A) as a function of its structural parameters.
Fig. \ref{fgr:fig3_uniaxialElongation}B shows the average mechanical response of fibrous scaffolds with random fiber arrangement as estimated using discrete fiber model and continuum finite element model. 
For small strain values, both the models demonstrate a nonlinear response of the scaffold (`J-shaped' curve, see \citep{Rizvi2012ActaBiomat}). 
Further, both of the models also showed good agreement with the experimental measurements on electrospun Polystyrene fibrous matrices for small strain values. 
Since we did not take fiber breakage into account in the models, at larger strains, the experimental measurements showed a deviation from the models. 
This disagreement at large strains is not of any major consequence  here since the focus of this work is on the mechanical response of the fibrous scaffolds against cellular forces which do not result in fiber breakage. 
\subsubsection{Sample size effect}
Owing to the fact that the ratio of the microscopic length scale $\ell$ and the sample size $L$ is not very small (Fig. \ref{fgr:fig2_schematic_fiberLenDist}B) for the fibrous scaffolds, the sample size is expected to influence its mechanical response in DFM. 
In order to study this effect we estimated the Young's modulus $\mathcal{Y}_m$ of the fiber scaffold with randomly arranged fibers for different sample dimensions (Fig. \ref{fgr:fig3_uniaxialElongation}C).
% For a fixed sample length $L$ (the dimension along the elongation direction) an increase in width $W$ results in an increase in sample's Young's modulus for small values of $W$ (Fig. \ref{fgr:uniaxialElon}C).
% However, for very large values of $W$ the Young's modulus attains a constant value.
% Further, it can be seen that the transition between these two behaviors takes place at smaller $W$ if the degree of fiber fusion is high (Fig. \ref{fgr:fig3_uniaxialElongation}D).
% We also estimated the Young's modulus for different sample sizes (Fig. \ref{fgr:uniaxialElon}D) and observed three types of behaviors. 
Here we observe mainly three types of behaviors.
For $L \gg W$, the Young's modulus of the fibrous scaffold sample is much smaller than that of the sample with $L \ll W$. 
For samples with aspect ratio close to $1$, that is $L \approx W$, the sample's Young's modulus is in between the two extremes. 
This effect of the sample size is more pronounced in the scaffold samples with low degree of fiber fusion and in sparse fibrous scaffolds (supplementary Fig. S2). 
We also observe the effect of sample size while changing the sample thickness by varying total fiber length $L_m$.
There too the Young's modulus of the scaffold increases with an increase in $L_m$ or sample thickness (Fig. \ref{fgr:fig3_uniaxialElongation}E).

The sample size effect on the elastic modulus can be intuitively explained for the fibrous scaffolds with no fiber fusion, that is $p_f=0$. 
For no fiber fusion, the resistance of the scaffold against uniaxial elongation can be written as the collective response of all the end-to-end fibers.
For random arrangement of linear elastic straight fibers, the force required for uniaxial elongation of the sample follows (see Appendix B for details)
\begin{equation}
    F_y (\varepsilon) \sim W \varepsilon \int \limits_0^W \int \limits_{x_1}^{x_2} \cos^3 \theta p_{\theta} (\theta) d\theta dx
\end{equation}
where $x_1=\tan^{-1}\left( -\cfrac{x}{L}\right)$ and $x_2=\tan^{-1}\left(\cfrac{W-x}{L}\right)$. 
For $W \ll L$, this simplifies to 
\begin{equation}
    F_y (\varepsilon) \sim \varepsilon W^2, \label{eq:Fy_eW2_smallW}
\end{equation}
and for $W \gg L$ we obtain
\begin{equation}
    F_y  (\varepsilon) \sim \varepsilon W. \label{eq:Fy_eW_largeW}
\end{equation}
This shows that for $W \ll L$ the Young's modulus of the fibrous scaffold increases with the sample width (Fig. \ref{fgr:fig3_uniaxialElongation}D).
However, for the sample size with $W \gg L$, Young's modulus attains a fixed value. 
% In such systems the mechanical response of the fibrous scaffolds under uniaxial elongation have been linked to the load bearing end-to-end fibers \citep{Rizvi2012ActaBiomat}.

In contrast to the DFM, the continuum FE model did not show any sample size effect (data not shown) on elastic modulus. 
This behavior is not surprising since continuum description of the scaffold assumes the microscopic length scale to be much smaller than the sample size.
% This is also reflected by the fact that a high degree of fiber fusion results in very weak sample size dependence on Young's modulus even in DFM (Fig. \ref{fgr:fig3_uniaxialElongation}D).
This demonstrates the limitation of continuum models for the scaffold mechanics where size of the cell is often comparable to microscopic length scale. 
\subsubsection{Effect of fiber fusion}
We also looked at the effect of the degree of fiber fusion on the mechanical response of the fibrous scaffolds against uniaxial elongation. 
For this we estimated the Young's modulus of fibrous scaffolds with different fiber fusion density. 
We find that an increase in the fiber fusion results in stiffer scaffold (Fig. \ref{fgr:fig3_uniaxialElongation}D, also reported in \citep{Razavi2020}). 
The fiber fusion also influences the sample size effect such that for higher fiber fusion the Young's modulus attains saturation value for smaller sample widths. 
This is an expected outcome since for higher fiber fusion, due to smaller $\ell$,  the ratio of microscopic and macroscopic length scale is very small.
% As shown in Fig. \ref{fgr:uniaxialElon}D the stiffness of the scaffold increases with the degree of  fiber alignment when the mean  fiber orientation coincides with the direction of the elongation. 
% However, for mean fiber orientation perpendicular to the elongation direction the scaffold stiffness demonstrates a decrease with the degree of fiber alignment.
% We have also found that an increase in the degree of fiber fusion results in a stiffer response of the scaffold under uniaxial elongation (see inset in Fig. \ref{fgr:uniaxialElon}). 
% This effect of fiber fusion has also been reported for uniaxial as well as biaxial elongational loading conditions \citep{Razavi2020}.  

These results on the matrix response against uniaxial elongation as a function of structural parameters, which can also be called `macro-mechanical response', have been studied extensively using experimentally as well as computational modeling.
This, however, does not shed any light on whether the cells interacting with the fibrous matrices also experience the same mechanics. %We will focus on this next. 
\subsection{Micro-mechanics of fibrous scaffold}
% Till now we have focused on the mechanical response of the fibrous scaffold under uniaxial elongation, which we can also call the macro-mechanical response of the scaffold.
% The forces applied by the cells to the scaffolds are of very different nature. 
The mechanical interaction of the cells with scaffold is via the focal adhesions \cite{Martino2018Review} which transfer localized microscopic cellular forces between the two. 
Therefore, in the following, we shift our attention to the scaffold micro-mechanics or its mechanical response against localized forces. 
\subsubsection{Scaffold deformation due to localized forces}
\begin{figure}[H]
    \centering
    \includegraphics[width=0.95\textwidth]{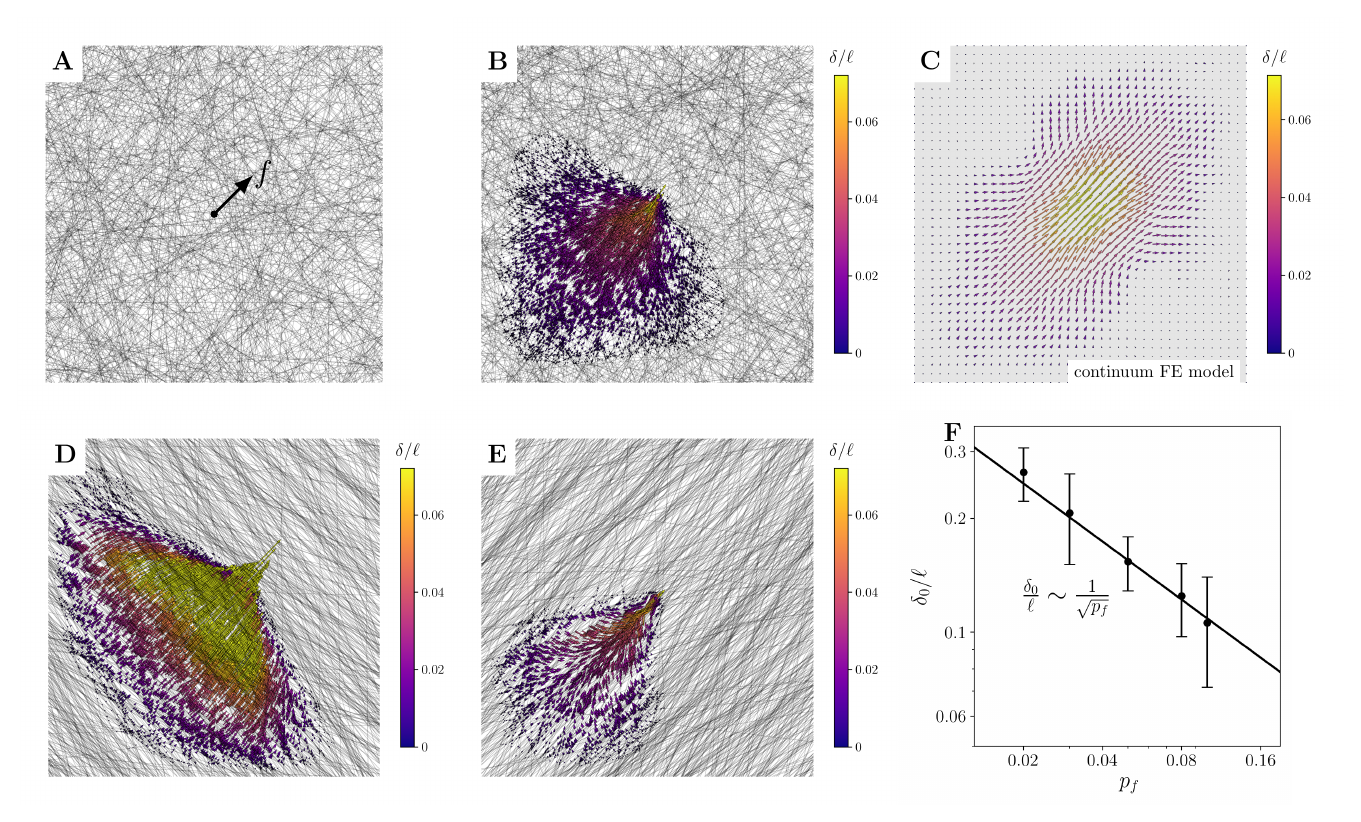}
    \caption{{\bf Fibrous scaffold micromechanics.} (A) Schematic showing application of localized point force. Displacement field in the fibrous scaffold with randomly arranged fibers as calculated from (B) discrete fiber model and (C) continuum finite element modeling. Displacement field in the fibrous scaffold with aligned fibers with point force (D) normal, and (E) parallel to the fiber alignment direction ($\theta_0=3\pi/4$ and $\pi/4$, respectively). (F) Displacement of the point at which force is applied as a function of fiber fusion density $p_f$ for random fiber arrangement. For (B), (D) and (E) $p_f=0.2$ and $L_m/W = 2\times 10^7$.}
    \label{fgr:fig4_pointForce}
\end{figure}

Towards this goal, we consider fibrous scaffold sample with all its boundaries fixed and apply a point force $f$ at its center (Fig. \ref{fgr:fig4_pointForce}A). 
This loading condition is somewhat similar to the well studied Kelvin problem in three dimensional elasticity \cite{PodioGuidugli2014}.
Fig. \ref{fgr:fig4_pointForce}B shows the displacement fields in the fibrous scaffolds due to a localized point force as estimated from DFM. 
Displacement field reveal anisotropic deformation of the fibrous scaffolds, even for the samples with random fiber arrangement. 
The anisotropic displacement field in case of the samples with random fiber arrangement can be attributed to the very small bending stiffness $k_b$ of the fibers as compared to their stretching stiffness $k_e$. 
In case of aligned fibers the scaffold deformation depends on the direction of the applied point force relative to the fiber orientation angle (Fig. \ref{fgr:fig4_pointForce}D-E). 
For a point force parallel to the fiber orientation the deformation field remains  localized to a smaller region as compared to when the directions of force and fibers are normal to each other. 
As opposed to DFM, the continuum FE model of the fibrous scaffold does not demonstrate anisotropy in the displacement field for the samples with random fiber arrangement (Fig. \ref{fgr:fig4_pointForce}C).
\subsubsection{Force transmission in fibrous scaffolds}
\begin{figure}[H]
    \centering
    \includegraphics[width=0.8\textwidth]{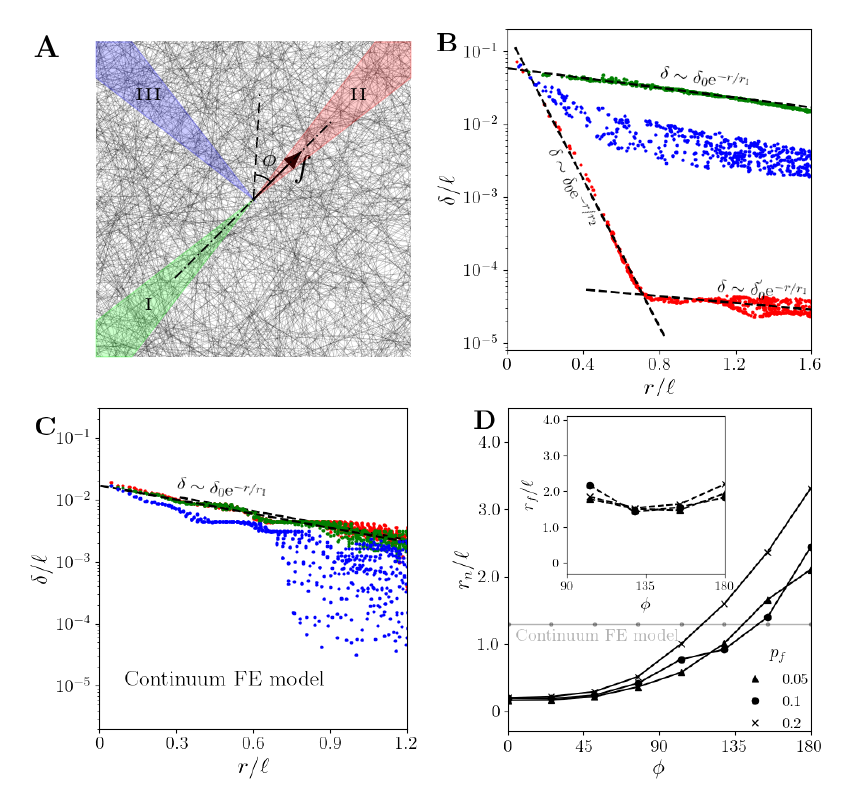}
    \caption{{\bf Displacement decay and characteristic length of force transmission.} (A) Schematic showing three regions relative to the applied point force $f$. Decay of displacement field as a function of $r$, the distance from the point force and, $\phi$, the position on the scaffold as obtained from (B) discrete fiber model ($p_f=0.2$) and (C) continuum FE model of scaffold with random fiber arrangement. (D) Characteristic decay lengths $r_n$ and $r_f$ for near and far field displacement, respectively, as a function of $\phi$ for different values of fiber fusion density $p_f$.  For $\phi<\pi/2$ the far field displacement is too small for $r_f$ calculation. In (B) and (D) $p_f=0.02$ and $L_m/W=2\times 10^7$.}
    \label{fgr:fig5_displacementDecay}
\end{figure}

For a quantitative insight into the nature of scaffold deformation we plotted the displacement field in the matrix as a function of the distance $r$ from the point of force application for a sample with randomly arranged fibers.
% In order to show the disparate nature of the deformation in different regions of the scaffold Fig. \ref{fgr:dispDecay}A shows the matrix displacement in three regions. 
For this, we divide the scaffold into multiple regions (Fig. \ref{fgr:fig5_displacementDecay}A) and the displacement plots for each region are shown in Fig \ref{fgr:fig5_displacementDecay}B (note the log scaling on $y$-axis). 
In region-I, the displacement field decays exponentially with $r$ and by fitting 
\begin{equation}
    \delta (r) = \delta_0 \exp \left( -r / r_1 \right)
\end{equation}
to the data we can estimate the characteristic decay length $r_1$ for region-I.
In contrast, for region-II, even though the displacement field shows exponential decay, it shows two different characteristic decay lengths- $r_2$ for points close to the point force ($r \ll 1$), and $r_1$ (same as that in region-I) for points far from the force ($r \gg 1$).
In region-III the scaffold response is somewhere in between that of the two extremes seen in regions-I and II. 
For region-II, the critical distance $r_c$ where transition between two behaviors takes place can be calculated as the intersection of two exponential decay curves.
We also observe that this critical transition distance does not depend very strongly on the degree of fiber fusion (supplementary Fig. S3).
This indicates that the anisotropy in the scaffold response is due to smaller bending stiffness of the fiber and not the fiber fusion.

We further plotted the near-field ($r_n$ for $r<\ell$) and far-field ($r_f$ for $r>\ell$) characteristic decay lengths as a function of angle $\phi$ from the direction of point force (Fig. \ref{fgr:fig5_displacementDecay}D). It shows that the far field decay length does not depend on $\phi$ indicating that the far-field displacement field is isotropic in nature. 
However, the near field decay length depends strongly on $\phi$. We observe that for small values of $\phi$ (region-II) $r_n \ll \ell$ whereas for $\phi \approx \pi$ (region-I) $r_n$ is large.
This indicates the anisotropic nature of the force transmission in the fibrous scaffold. 
The fibers under tension (in region-I) transmit force to larger distance than the fibers under compression in region-II.

Even though the exponential decay of the displacement field is seen in the continuum FE model as well (Fig. \ref{fgr:fig5_displacementDecay}C) the anisotropy of the response is not observed (Fig. \ref{fgr:fig5_displacementDecay}D). 
In fact, in the continuum model we obtain the characteristic decay length which does not depend on the direction.
Even though there is no experimental data available for the scaffold deformation due to in-plane localized forces (as far as we know), the displacement field due to a polarized cell (see inset in Fig. \ref{fgr:fig6_dipole}B) shows the anisotropic nature of scaffold deformation.
This shows that the continuum model predicts an isotropic force transmission in the fibrous scaffolds with random fibers.
% This suggests the limitation of the continuum FE models in predicting the scaffold mechanical behavior against microscopic loading conditions.
\subsubsection{Microscopic scaffold stiffness}
We can quantify the stiffness of the fibrous scaffold against the localized force as $E_{\mu} = (f/k_e) /(\delta_0/L)$, the ratio of the magnitude of the force applied at the center and the displacement of that point \cite{beroz2017physical}. 
Therefore, for a fixed amount of applied force microscopic stiffness of the scaffold can be characterized by the displacement of the center point with stiffer scaffolds showing smaller displacements.
Fig. \ref{fgr:fig4_pointForce}F shows the displacement of center point $\delta_0$ as a function of the degree of fiber fusion $p_f$ for a fixed value of applied force. 
This shows an increase in the microscopic stiffness of the scaffold with increasing fiber fusion.
Not only that, the numerical fit to the simulation data also reveals that the displacement of the center point scales as $\cfrac{\delta_0}{\ell} \sim \cfrac{1}{\sqrt{p_f}}$ for small values of $p_f$.
It has to be pointed out that this scaling might not hold for larger values of $p_f$ since that would imply a rigid nature of scaffold for large $p_f$. 
This increase in microscopic stiffness with increasing fiber fusion also corresponds to the increase in the macroscopic stiffness (Fig. \ref{fgr:fig3_uniaxialElongation}D).
This estimate and scaling law gives us an insight which can be utilized in modulating the fiber fusion in the fibrous scaffolds for desired tissue engineering applications. 
% For application of force parallel to the fiber alignment direction, the scaffold deformation is very restricted.
% On the other hand, for force direction normal to the fiber alignment direction the scaffold deformation 
% Fiber alignment, Force displacement as a function of degree of fiber fusion. 
\subsection{Stiffness sensing of a fibrous scaffold by a cell}
\begin{figure}[H]
    \centering
    \includegraphics[width=0.8
    \textwidth]{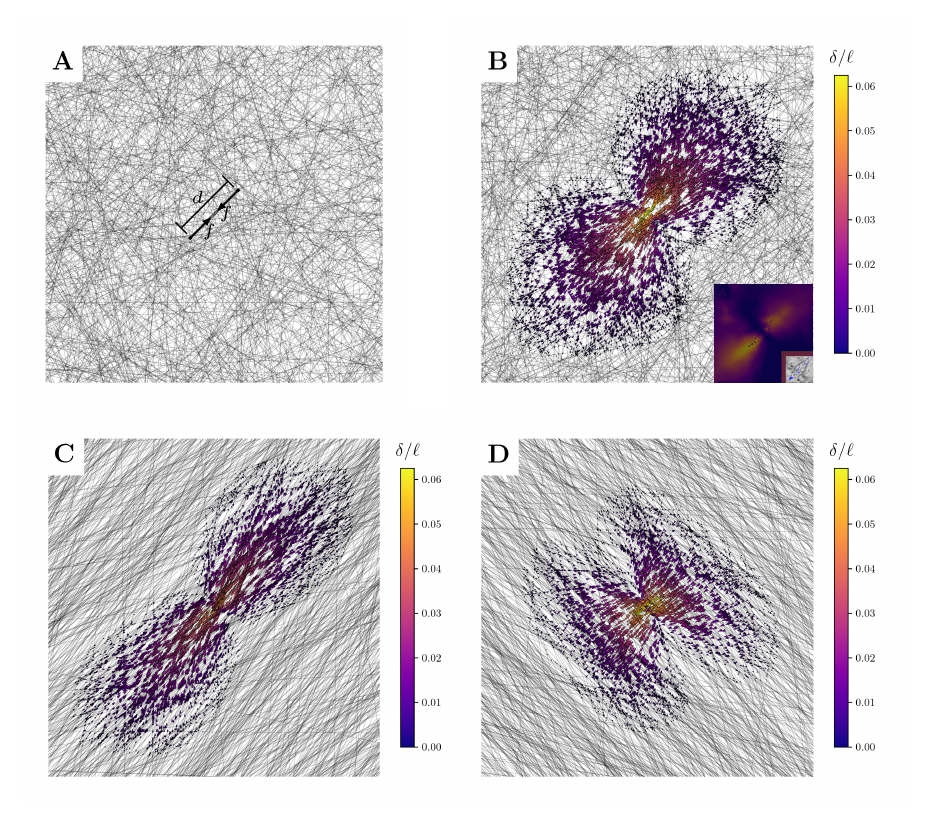}
    \caption{{\bf Scaffold deformation due to a cellular force dipole.} (A) Schematic showing placement of cellular force dipole on fibrous scaffold. (B) Displacement field due to  force dipole in fibrous scaffold with random fiber arrangement. Inset shows the displacement field due to an epithelial cell on a collagen scaffold (adapted from \cite{gjorevski2012}, with permission). Displacement field in fibrous scaffold with aligned fibers when the dipole is placed (C) parallel and (D) normal to the fiber orientations. Here $p_f=0.2$, $L_m/W=2\times 10^7$ for (B-D), and $\theta_0=\pi/4$ and $3\pi/4$ for (C) and (D), respectively.}
    \label{fgr:fig6_dipole}
\end{figure}
As mentioned earlier cells apply localized forces to the underlying substrate through the focal adhesion contacts to sense the mechanical properties of the scaffold \cite{Janmey2020Review}. 
Despite the extensive works on the biophysics of this process the exact mechanism of the stiffness sensing by the cells has remain elusive \cite{DeSafran2008}.
Since the exact measurement of the force acting on each focal adhesion is not straight forward, the forces applied by the cells are usually modeled as force dipoles \cite{beroz2017physical,DeSafran2008}. 
In its general form, a force dipole is defined as the second order tensor
\begin{equation}
    \underline{\underline{\mathcal{P}}} = \int \mathbf{r} \otimes \mathbf{t} dl 
\end{equation}
where $\mathbf{r}$ and $\mathbf{t}$ are the position vectors of the focal adhesion and the localized force acting on them. 
Here the integration is performed along the boundary of the cell.
Since the cells usually do not apply any toque to the substrate the force-dipole tensor is symmetric in nature with its principal directions (corresponding to the eigenvectors of the matrix representation) coinciding with the directions of maximum and minimum forces applied by the cells.
Assuming the force in one principal direction to be much larger as compared to that in the other, the force-dipole model of a cell reduces down to two opposing point forces $f$ separated by a distance $d$ (Fig. \ref{fgr:fig6_dipole}A).
We looked at the deformation of the fibrous scaffold due to this force-dipole to understand the cell-scaffold interaction and stiffness sensing by a cell.
\subsubsection{Scaffold deformation due to a cell}
Fig. \ref{fgr:fig6_dipole}B shows the displacement field of the fibrous scaffold with random fiber arrangement due to the forces applied by a force dipole (as shown in Fig. \ref{fgr:fig6_dipole}A). 
Here too, the displacement field shows anisotropy but that is due to the nature of the forces applied by the force-dipole. 
This displacement field is in qualitative agreement with the experimental measurements for an epithelial cell on a collagen scaffold (inset Fig. \ref{fgr:fig6_dipole}B).
On the scaffold with aligned fibers the displacement field shows dependence on the cell orientation relative to fibers (Fig. \ref{fgr:fig6_dipole}C-D). 
Apart from the qualitative difference between the two aligned systems (Fig. \ref{fgr:fig6_dipole}C and Fig. \ref{fgr:fig6_dipole}D) the magnitude of the displacement (and thereby scaffold deformation) is smaller when the force dipole orientation is parallel to the fibers.
This is due to primarily stretching and bending deformations of the fibers in Fig. \ref{fgr:fig6_dipole}C and Fig. \ref{fgr:fig6_dipole}D, respectively.
\subsubsection{Scaffold stiffness sensing by a cell}
It is still a matter of debate \cite{Panzetta2019PNAS} whether cells apply a fixed amount of force to the substrate and sense stiffness based on the magnitude of the substrate deformation or they apply a fixed amount of strain and stiffness is sensed in terms of the resistance force offered by the substrate.
Here we test both of these possibilities to identify if the stiffness sensed by a cell depends on cellular mechanism and how does stiffness sensed by the cell depend on the cell's own stiffness. 

In both cases, in principle, the cell can sense the scaffold stiffness in terms of the difference between its deformation with and without substrate interaction. 
Therefore, for small magnitude of cell deformation the scaffold stiffness sensed by the cell is  
\begin{equation} \label{eq:siffnessSensed}
    \mathcal{E}_m = \frac{\sigma_0}{\epsilon_c} - \mathcal{E}_c
\end{equation}
where $\sigma_0 = f/A_{\text{FA}}$ is the stress defined as the ratio of the force by the cell and the cross-sectional area of a focal adhesion, %(focal adhesion area - 0.25-10um depending on the maturation state of FA)
$\epsilon_c$ is the strain in the cell when it is attached to the scaffold.
Here the cell stiffness $\mathcal{E}_c$ is an effective parameter which takes into account the contributions from the cytoskeleton, focal adhesion, and other cellular components.  
For the scenario when cell applies a fixed amount of force (force-dipole) $f$ the strain in the cell $\epsilon_c$ is the unknown which we have to calculate. 
On the other hand, when the fixed strain is applied by the cell (strain-dipole), we have to estimate the force $f$ required by the cell. 
We performed numerical simulations for fiber scaffolds with random fiber arrangements and force dipole as a model of cell in both of these conditions. 

\begin{figure}[H]
    \centering
    \includegraphics[width=0.73\textwidth]{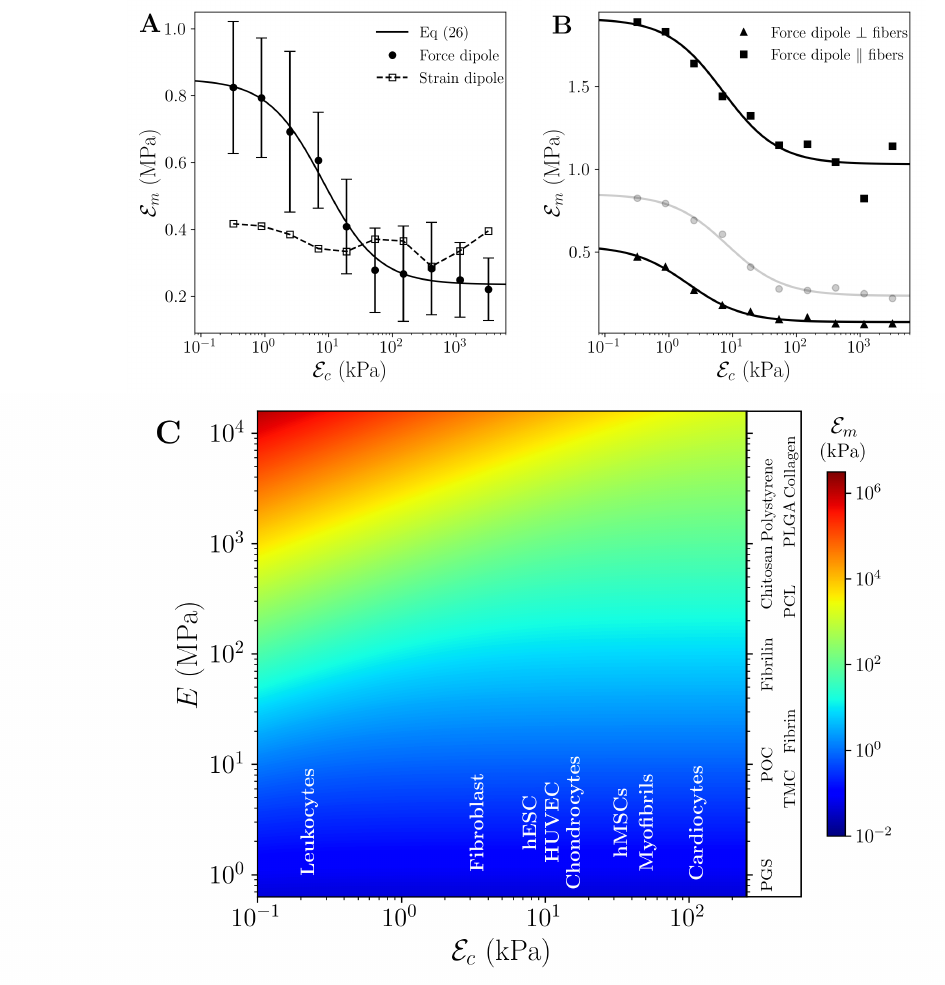}
    \caption{{\bf Stiffness sensing of fibrous scaffold by a cell.} 
    (A) Dependence of stiffness of the fibrous scaffold with random fiber arrangement, as sensed by a cell by applying constant force (force dipole) and constant deformation (strain dipole), on the cell stiffness. The error bars in the strain dipole case are not shown for clarity. 
    (B) Stiffness sensing of fibrous scaffold with aligned fibers by a cell via force dipole. The gray curve for the scaffold with random fiber arrangement (same as in panel A) is also shown for reference. 
    (C) Estimated stiffness sensing of fibrous scaffolds (with randomly arranged fibers) composed of materials with different elastic properties by different cell types. In (C) $E$ is the Young's modulus of the bulk material of the fibers. The exact values of the elastic properties of materials and cells shown in (C) are listed in supplementary information. For all panels $p_f=0.02$ and $L_m/W=2\times 10^7$.}
    \label{fgr:fig7_stiffnessSensing}
\end{figure}

Fig. \ref{fgr:fig7_stiffnessSensing}A shows the stiffness of the fibrous scaffold with random fiber arrangement as sensed by the cell as a function of the cell stiffness. 
It can be seen that the scaffold stiffness sensed by the cell does not depend on the cell stiffness when the cell controls the deformation and not force (dashed curve in Fig. \ref{fgr:fig7_stiffnessSensing}A).
However, when cell applies a constant amount of force the stiffness sensed by the cell decreases with increasing cell stiffness.

In order to understand this interesting behavior we looked at this more closely using a simplified lumped-parameter model of stiffness sensing. 
Such models have previously been used to study the cell-substrate interactions \citep{DeSantis2011}.
In the lumped-parameter model we consider stiffness sensing by a cell in one dimension where a cell (elastic modulus $\mathcal{E}_c$, length $d$) is attached to a 1D scaffold (length $2l+d$) on two sides and the two ends of the substrate are held fixed.
We consider that the stress-strain relation for this 1D scaffold is nonlinear and is given by function $\sigma_m (\epsilon_m)$ (with $ \left. \cfrac{d \sigma_m}{d \epsilon_m} \right|_{\epsilon_m \rightarrow 0} = \mathcal{Y}_m$) where $\epsilon_m$ is the strain in the scaffold.
For $\epsilon_c$ strain in the cell the strain in the scaffold on each side of the cell is $\epsilon_m=\cfrac{\epsilon_c d}{2l}$. 
The stress applied by the cell to generate this deformation is, therefore, given by
\begin{equation} \label{eq:cellForceScaffold}
    \sigma_0 = \mathcal{E}_c \epsilon_c + \sigma_m \left( \frac{\epsilon_c d }{2l}\right).
\end{equation}

Therefore when cell applies a constant amount of strain the stiffness of the scaffold sensed by the cell can be calculated from equation \eqref{eq:siffnessSensed}  and is given by 
\begin{equation}
    \mathcal{E}_m = \left. \frac{1}{\epsilon_c} \sigma_m \left( \frac{\epsilon_c d }{2l}\right) \right|_{\epsilon_c \rightarrow 0} \sim \mathcal{Y}_m.
\end{equation}
This shows that in the strain controlled scenario the stiffness sensed by the cell is proportional to the macroscopic stiffness and does not depend on the stiffness of the cell. This is in agreement with the numerical simulation of the scaffold model (Fig. \ref{fgr:fig7_stiffnessSensing}A).

On the other hand, when cell applies a constant force we need to solve equation \eqref{eq:cellForceScaffold} for $\epsilon_c$ which is unknown in this case. 
This will require the knowledge of the exact form of the constitutive relation of the scaffold.
For our purposes, since the fibrous scaffold shows nonlinear relation between stress and strain (Fig. \ref{fgr:fig3_uniaxialElongation}B), we consider a small degree of nonlinearity in the scaffold constitutive model and assume that it has the form
\begin{equation}
    \sigma_m(\epsilon_m) = \mathcal{Y}_m \epsilon_m + \alpha \epsilon_m^2 
\end{equation}
where $\alpha$ is a small phenomenological parameter. 
We have chosen a small degree of nonlinearity ($\alpha \ll 1$) to keep the analysis tractable. 
In principle, the force-displacement curve due to the localized force can be utilized for more accurate calculation. 
Following a perturbative approach, we can now solve equation \eqref{eq:cellForceScaffold} to obtain
\begin{equation}
    \epsilon_c = \frac{\sigma_0}{\mathcal{S}}  \left( 1 - \frac{\alpha \sigma_0}{\mathcal{S}^2  } \right)
\end{equation}
where $\mathcal{S} = \mathcal{E}_c + \mathcal{Y}_m \cfrac{d}{2l}$. We can substitute this in equation \eqref{eq:siffnessSensed} to obtain
\begin{equation}
    \mathcal{E}_m = %\mathcal{Y}_m \frac{d}{2l} + \frac{\alpha}{\mathcal{S}} \sigma_0 = 
    \mathcal{Y}_m \frac{d}{2l} + \frac{\alpha}{\left(\mathcal{E}_c + \mathcal{Y}_m \cfrac{d}{2l}\right)} \sigma_0
\end{equation}
which shows the dependence of scaffold stiffness sensed by the cell on $\mathcal{E}_c$. 
Fig. \ref{fgr:fig7_stiffnessSensing}A-B shows that this relation fit very well with the numerical results obtained for the fibrous scaffold. 

The above analysis also demonstrates that if $\alpha = 0$, that is scaffold response is linear elastic, then the scaffold stiffness sensed by the cell is independent of the cell stiffness and also does not depend on the details of the cellular machinery (constant strain vs constant stress).
This dependence of the cell sensed stiffness on $\alpha$, the non-linearity in the scaffold response, can be intuitively understood in the following manner.
Considering the forces applied by the cells to be same the deformation in a stiffer cell (a cell with high $\mathcal{E}_c$) is smaller.
This results in smaller deformation of the scaffold as well.
Owing to the non-linearity in the scaffold constitutive response we get smaller $\mathcal{E}_m$ for smaller deformations, resulting in decrease in $\mathcal{E}_m$ with increasing $\mathcal{E}_c$.
This mechanism is further visible when we consider strain applied by the cell instead of stress as the model for the cell forces (Fig. \ref{fgr:fig7_stiffnessSensing}A).
In this scenario, the deformation in the scaffold is not dependent on cell stiffness and therefore the resultant $\mathcal{E}_m$ is independent of $\mathcal{E}_c$. 

For the scaffold with aligned fibers also we find similar behavior of cell stiffness dependent sensing of the scaffold stiffness (Fig. \ref{fgr:fig7_stiffnessSensing}B). 
For scaffold with aligned fibers the orientation of the cell relative to the fiber direction influences the stiffness sensed by the cell.
This results in higher and lower values of $\mathcal{E}_m$ for cell oriented parallel and normal to the fiber direction, respectively.
This effect is due to the fact that for a force-dipole oriented parallel to the fibers the scaffold deformation is mainly due to the stretching of the fibers as opposed to the fiber bending, which offers lesser resistance as compared to fiber stretching, when force-dipole is orientation normal to the fibers. 

From this insight into the effect of cell stiffness on scaffold stiffness sensing via force-dipole we created a map of combinations of fibrous scaffolds of different materials and their stiffness sensing by several cell types (Fig. \ref{fgr:fig7_stiffnessSensing}C).
This maps shows that the cells such as leukocytes (with stiffness of the order of $0.07-0.2$kPa \cite{rosenbluth2006force}) and cardiocytes (with stiffness of the order of $100$kPa \cite{mathur2001endothelial}) will sense difference stiffness of the same fibrous scaffold. 
As shown in the figure that this cell type dependence on the sensed stiffness can be very large upto multiple orders of magnitude. 
Further, the effect of cell type is also dependent on the scaffold material itself. 
For an example, for fibrous scaffold composed of relatively soft materials (such as fibrin with Young's modulus $\approx 15$MPa \cite{collet2005elasticity}) the variation in the sensed stiffness by different cells is very small (see Fig. S4). 
On the other hand, for scaffold made up of stiffer materials (for example polystyrene with Young's modulus $\approx 3$GPa \cite{oral2011measuring}) this variation is multiple orders of magnitude (Fig. S4).
This demonstrates a quite intricate relationship between the cell type and scaffold material on the stiffness sensing by the cells.
% \subsubsection{Effect of fiber alignment and fusion}
\section{Discussion}
Mechanical properties of tissue engineering scaffolds have been known to be critical for their application in tissue regeneration applications. 
This has propelled the research on understanding their mechanics experimentally as well as by mathematical and computational means, extensively.
This has resulted in extensive works in this direction but most of the insights into the mechanical behavior of biomaterials in general and fibrous scaffolds in particular have been from the macroscopic view point \cite{sacks2000biaxial,ghezelbash2021shear}. %buffinton2015uniaxialElongation, chaudhry2009viscoelastic
The understanding of the mechanical response of these materials at the length scale of a single cell has been quite limited.
The present work is an attempt to fill this gap using a computational model of fibrous scaffold mechanics.

% \subsection{Sample size effects on fibrous scaffold mechanics}
A consideration of the microscopic length scale of scaffold ($\ell$ in this work) not very small as compared to the sample size or the size of a cell results in some very interesting properties. 
For example, we observe the effect of the sample size on the mechanical response of scaffold under uniaxial elongation for $\ell \not\ll L,W$. 
Even though this sample size effect is very well known in other materials with a well defined microstructure \cite{zhu2008materials} its effect on fibrous biomaterials has not been very well studied \citep{Rizvi2016BMM}. %Rizvi2012ActaBiomat
It illustrates that during experimental testing of fibrous scaffolds under uniaxial or biaxial loading conditions a care is warranted. 
It is required that during these mechanical testing conditions materials are tested with multiple sample sizes so that the effects of microscopic length scale (if any) can also be taken into account during elastic moduli calculations. 
For instance for fibrous scaffolds with random fiber arrangement sample with large width should be used in uniaxial elongation testing for a consistent measurement of the Young's modulus (Fig. \ref{fgr:fig3_uniaxialElongation}D-E).

Further, the sample size effect can also be used for indirect estimation of the extent of fiber fusion in the scaffolds. 
Since in scaffolds with higher fiber fusion the effect of sample width is smaller (Fig. \ref{fgr:fig3_uniaxialElongation}D) we can estimate the microscopic length by performing uniaxial mechanical testing with different sample widths. 
This information can then be used with Fig. \ref{fgr:fig2_schematic_fiberLenDist}B to quantify the degree of fiber fusion in scaffold. 

This effect of the microscopic length scale, however, does not reflect in the finite element analysis of the continuum homogenization based models \cite{HolzapfelGasserOgden2000} of the fibrous scaffold under same condition as that in DFM. 
This is not surprising since the formulation of the continuum model (such as HGO model used in this work) is based on the assumption that the microscopic length scale is very small as compared to the sample size. 
This also means that for the situations where the cell size is comparable to the microscopic length, which is often the case in tissue engineering application, the utilization of continuum modeling is not completely justifiable. 
Another important advantage of the discrete fiber modeling over continuum approach is due to the fact that from DFM one can extract the mechanical response of a particular scaffold sample with a known (possibly from a non-destructive imaging technique such as computed tomography) fiber architecture. 
Continuum models of fibrous scaffolds, by virtue of dealing with average mechanical quantities, usually do not provide information about the variability in the scaffold response which becomes important when the microscopic length scale is not too small. 

% \subsection{Scaffold macro-mechanics versus micro-mechanics}
The dependence of the sensitivity of the stiffness sensing by the cell on the cell's own elastic properties presents another challenge to the practice of macroscopic mechanical testing (uniaxial/biaxial elongation or shear deformation) for the characterization of scaffold mechanics for tissue engineering.
The use of macroscopic elastic moduli as a measure of the mechanical behavior of the scaffold are justifiable as long as the force/deformation applied by the  cell is small enough where the constitutive relation of the substrate can be linear.
The linear constitutive relation, even for small deformations, however, is not always correct for the fibrous materials \citep{Rizvi2014JMBBM}.
It is possible to have two structurally very different fibrous scaffolds which can show identical stress-strain curves under uniaxial elongation tests but provide completely different micro-mechanical environment to the cells \citep{Rizvi2016BMM}. 
This limitation of the macroscopic mechanical tests has been acknowledged and has resulted in the use of atomic force microscopy (AFM) for the estimation of local mechanical properties of the biomaterials \cite{iturri2017afm}. %dutta2015nano
This, however, does not resolve the issue completely since the nature of the mechanical forces applied by the AFM (usually normal to the substrate surface) and the forces applied by the cells (tangential to the substrate surface) are not the same.
Therefore, until a new experimental technique is devised which can assess scaffold mechanics under loading condition similar to that by a cell one has to apply combine mathematical modeling with the experimental measurements for assessing the micro-mechanical environment experienced by the cells. 

% \subsection{Effect of cell type on cell-scaffold interactions}
One of the many functions of the force application by the cells is also the probing the mechanical properties of the underlying substrates \cite{Janmey2020Review,discher2005tissue}.
The debate on the exact mechanism of stiffness sensing by the cells is still not settled \cite{DeSafran2008,Panzetta2019PNAS}.
As mentioned previously, it, however, has been distilled to a handful of mechanisms involving sensing of resistance force by the cell
or sensing of  deformation 
or, more recently reported sensing of the stored energy \cite{Panzetta2019PNAS}. 
In this work we have estimated the fibrous scaffold's  stiffness as a function of cell's own elasticity which shows that the deformation sensing by the cell (after application of a fixed amount of force) can show cell type dependent response where the difference between the stiffness of a scaffold sensed by a stiff cell (such as a cardiocyte) and that by a soft cell (such as a leukocyte) can be as high as an order of magnitude (Fig. \ref{fgr:fig7_stiffnessSensing}C). 
Further, we have also seen that the dependence of stiffness  sensing (by force-dipole) on cell stiffness can be attributed to the smaller overall deformation of the system by stiffer cells.
This also implies that the stiffness sensed by the cells may also depend on the magnitude of the cell forces where a cell applying small force will experience a softer scaffold. 
This can be an important factor since all the cells do not apply same amount of force \cite{pelham1999high,burton1999keratocytes,yin2005myocite}.%lee1994traction
The force sensing by the cell after application of a fixed amount of deformation, however, does not depend on the mechanical properties of cell.
Even though these observations require experimental validation they also suggest possible avenues to look into to understand the mechanisms of mechanosensing. 
It is also not unlikely that different cell types use different mechanisms for the same goal of mechanosensing and, therefore, we may not have a single `winner' among different possibilities.

% In this work we have looked at the dependence of mechanosensing on cell stiffness but the sensitivity of stiffness sensing on substrate stiffness has been studied extensively [REF]. 
% These works have shown that .... 
Furthermore, here we have assumed that the cell applies a fixed amount of force irrespective of the scaffold stiffness. 
This may appear to be in conflict with the fact that cells undergo cytoskeletal remodeling \cite{Janmey2020Review,Janmey2009Review} %jannatbabaei2019cytoskeletal
depending on the scaffold stiffness and, therefore, the force/deformation applied by the cells can depend on the scaffold stiffness.
Therefore, it needs to be pointed out here that the mechanosensing we are studying in this work is only limited to the very early stages of the cell contact with the scaffold when the cytoskeletal remodeling has not yet taken place.
Of course, in order to look at the long term response of the cell as a function of scaffold stiffness we have to take into account the cytoskeletal remodeling and focal adhesion dynamics \cite{gupta2016single}. %fusco2017mechanosensing 
% \begin{enumerate}
    % \item Different cell types may sense different stiffness of the same scaffold. Proposed experiment ?
    % \item Numbers - stiffness of typical materials used in TE. stiffness of different cell types. 
% \end{enumerate}

% \subsection{Model limitations and future scope}
Finally, it is also important to keep an eye on the limitations of the discrete fiber model presented here. 
First, the consideration of the scaffold to be a two dimensional structure puts a limitation on its applicability. 
This assumption, however, is justified in case of the electrospun fibrous scaffolds where the sample thickness is almost always very small as compared to its other dimensions \cite{ryu2020thickness}.

Furthermore, here we take into account the fusion between the fibers (also studied in few other works \citep{Razavi2020}) but fusion points are modeled only to transfer mechanical fibers among the participating fibers. 
This simplification is justifiable for the fibrous scaffolds where the fiber deformation is stretching dominated which is the case in this work.
For the `soft' fibrous scaffolds with highly curved fibers (that is $\xi \gg 1$) the deformation is mainly fiber bending and there we have to also take into account the deformation of the fiber-fiber junctions.
Here we have also assumed the fiber fusion to be perfect, that is the fused fibers do not detach from each other, which may not be the case when large deformations are imposed on the scaffolds. 
The microscopic cellular forces, however, are not usually large enough to cause breaking of fibers or fiber junctions.
Furthermore, in addition to the fiber fusion the mechanical interaction among fibers can also be due to the friction between \cite{barbier2009friction} the two fibers in contact which has also been ignored here. 

Here, in order to keep the system as simple as possible we modeled the cellular forces by a force or strain dipole. 
Even with this simple representation the displacement filed obtained has a qualitative similarity with that in  experiments \cite{gjorevski2012}.
For a more detailed understanding we also need to model the cell with more details where cell geometry, ligand-receptor based cell-scaffold adhesion, cytoskeletal remodeling and focal adhesion dynamics are also taken into account. 
Among these factors the focal adhesion dynamics as a function of the substrate stiffness is very important since it also governs the short and long term response of the cell on fibrous scaffold. 
Similarly, for fibers also more detailed Euler-Bernoulli \cite{bauchau2009euler}
or Timoshenko \cite{ochsnerclassical}
beam based models should be taken into account if the cell size becomes comparable to that the fiber dimensions. 

\section{Conclusions}
In this work we have looked at the mechanics of the tissue engineering fibrous scaffolds using mathematical modeling.
Modeling of the mechanical behavior of fibrous scaffolds present a unique challenge since their microscopic length scale is usually comparable to physical length or size of the cells. 
We have explained that this property results in sample size dependent macroscopic mechanical properties.
This work also presents the microscopic mechanics of fibrous scaffold due to localized cellular forces which generate anisotropic deformation which decays exponentially with distance. 
Finally, the mathematical model is also used to study the mechanism of stiffness sensing by the cells. 
This shows that the stiffness of the scaffold sensed by the cell can depend very strongly on the cellular machinery (force dipole vs strain dipole) and cell stiffness. 
These demonstrations will help not only in designing of tissue engineering scaffolds with specific application but also in formulating the experimental strategies for the study of the mechanosensing by the cells. 
\section*{Appendix A: Estimation of average fiber curvature}
\setcounter{equation}{0}
\renewcommand{\theequation}{A.\arabic{equation}}
Curvature of a fiber (considering it as a two dimensional curve for 2D DFM) is a characteristic property which is dependent on its exact shape and is usually defined at each location along the fiber.
In the discrete fiber model, however, we consider only a statistical description of the fibers in terms of their normalized length $\xi=l/R$ (a single scalar quantity for each fiber) and do not take into account the exact shape of each fiber (Fig. \ref{fgr:fig_AppendixA}).
Therefore, for the constitutive modeling of the bending of the fibers we have to estimate the average fiber curvature in terms of $\xi$. 
\setcounter{figure}{0}
\renewcommand{\thefigure}{A.\arabic{figure}}
\begin{figure}[H]
    \centering
    \includegraphics[width=0.35 
    \textwidth]{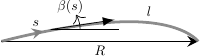}
    \caption{Schematic showing average curvature estimation for a fiber.}
    \label{fgr:fig_AppendixA}
\end{figure}
As shown in Fig. \ref{fgr:fig_AppendixA} the exact geometry of fiber in 2D can be described in terms of the tangent angle $\beta(s)$ made by the fiber with its director vector (vector joining its two ends). 
This angle is a function of arclength parameter $s$ defined along the fiber length.
Therefore we have 
\begin{equation}
    \frac{1}{\xi} = \frac{R}{l} = \int \limits_0^1 \cos(\beta(s)) ds. \label{eq:oneOverXi}
\end{equation}
In order to approximate the shape of the curved fiber we assume that the tangent angle $\beta$ is given by $\beta(\tilde{s}) = \sum \limits_{n=1}^{\infty} \beta_n = \sum \limits_{n=1}^{\infty} a_n \sin(n \pi \tilde{s})$ where $n$ is an integer and $\tilde{s}=s/l$ is the normalized arclength parameter.  
Therefore, the amplitude of the $n^{\text{th}}$ mode $a_n$ can be obtained from the equation \eqref{eq:oneOverXi}.
If we assume that the fibers are not too curved, that is $\xi \not\gg 1$, we can approximate the fiber shape by only first mode to obtain
\begin{equation}
    a_1 = 2\sqrt{2} \sqrt{1-\frac{1}{\sqrt{\xi}}}.
\end{equation}
Therefore, the average root mean square curvature of the fiber can be written as 
\begin{equation}
    \langle \kappa_0 (\xi) \rangle = \left(\int \limits_0^1 \left( \frac{d\beta}{d\tilde{s}}\right)^2 d\tilde{s} \right)^{1/2} \approx 2 \pi \sqrt{1-\frac{1}{\sqrt{\xi}}}.
\end{equation}
\section*{Appendix B: Sample size effect on Young's modulus}
\setcounter{equation}{0}
\renewcommand{\theequation}{B.\arabic{equation}}
For a rectangular sample of dimensions $L \times W$ (Fig. \ref{fgr:fig_AppendixB}) with straight fibers, where fiber orientation
is given by probability density function $\ptheta (\theta)$, the strain in a fiber oriented at angle $\theta$ when $\varepsilon \ll 1$ elongation strain is
applied on the sample is given by
\begin{equation}
    \varepsilon_f = \frac{L\sqrt{\tan^2\theta + (1 + \varepsilon)^2} -L\sqrt{\tan^2\theta + 1} }{L\sqrt{\tan^2\theta + 1}} \approx \varepsilon \cos^2 \theta
\end{equation}
If the fiber material is linear elastic with Young's modulys $Y_f$ and cross-sectional are $A_f$ then the resistance force due to this one fiber in the direction of elongation is 
\begin{equation}
    f_y (\theta) = Y_f \varepsilon_f A_f \cos \theta \approx Y_f A_f \varepsilon \cos^3 \theta.
\end{equation}
Writing the resistance of the scaffold as the collective response of all the fibers, we get 
\begin{equation}
    F_y = N_0 \int \limits_0^W \left( \int \limits_{x_1}^{x_2} f_y (\theta) \ptheta (\theta) d \theta \right) dx,
\end{equation}
where $N_0$ is the number of end-to-end fibers per unit width of the sample. 
For $W \ll L$, we have $x_1 \approx -x/L$, $x_2 \approx (W-x)/L$ and $\cos \theta \approx 1$ which give
\begin{equation}
    F_y \approx N_0 Y_f A_f \varepsilon \int \limits_0^W \int \limits_{-x/L}^{(W-x)/L} \ptheta(0) d\theta dx = \frac{N_0Y_fA_f\varepsilon \ptheta(0) W^2}{L}.
\end{equation}
On the other hand, for $W \gg L$ with uniform probability density $\ptheta(\theta)=1/\pi$ we obtain
\begin{equation}
    F_y \approx \frac{N_0 Y_f A_f \varepsilon}{\pi} \int \limits_0^W \int \limits_{x_1}^{x_2}\cos^3 \theta d\theta dx = \frac{4}{3} \frac{N_0 Y_f A_f \varepsilon W}{\pi}.
\end{equation}
Calculating Young's modulus of the scaffold as $\mathcal{Y}_m = \cfrac{F_y}{W \varepsilon t}$,   with $t$ being sample thickness, we obtain the dependence of sample width on Young's modulus.
\setcounter{figure}{0}
\renewcommand{\thefigure}{B.\arabic{figure}}
\begin{figure}[H]
    \centering
    \includegraphics[width=0.25\textwidth]{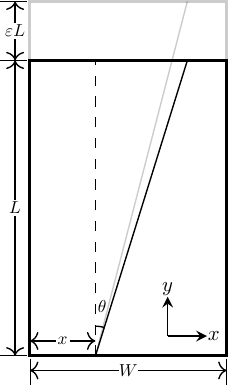}
    \caption{Schematic showing uniaxial elongation of a rectangular fibrous scaffold sample.}
    \label{fgr:fig_AppendixB}
\end{figure}
% \end{multicols}

\section*{Acknowledgements}
We are grateful to Dr. Prasoon Kumar for providing electron micrographs of electrospun scaffolds for Fig. \ref{fgr:schematic}. 
DJM thanks the MoE of scholarship and 
MSR was supported by Institute Seed Grant from Indian Institute of Technology Hyderabad. 
% \nocite{*}
\bibliographystyle{unsrt}
\bibliography{forArxiv}

\begin{thebibliography}{10}

\bibitem{Ringe2012}
J.~Ringe, G.~R. Burmester, and M.~Sittinger.
\newblock {{R}egenerative medicine in rheumatic disease-progress in tissue
  engineering}.
\newblock {\em Nat Rev Rheumatol}, 8(8):493--498, 08 2012.

\bibitem{Langer2016}
A.~Khademhosseini and R.~Langer.
\newblock {{A} decade of progress in tissue engineering}.
\newblock {\em Nat Protoc}, 11(10):1775--1781, Oct 2016.

\bibitem{Rustad2010}
K.~C. Rustad, M.~Sorkin, B.~Levi, M.~T. Longaker, and G.~C. Gurtner.
\newblock {{S}trategies for organ level tissue engineering}.
\newblock {\em Organogenesis}, 6(3):151--157, 2010.

\bibitem{Howard2008}
D.~Howard, L.~D. Buttery, K.~M. Shakesheff, and S.~J. Roberts.
\newblock {{T}issue engineering: strategies, stem cells and scaffolds}.
\newblock {\em J Anat}, 213(1):66--72, Jul 2008.

\bibitem{Salinas2018}
E.~Y. Salinas, J.~C. Hu, and K.~Athanasiou.
\newblock {{A} {G}uide for {U}sing {M}echanical {S}timulation to {E}nhance
  {T}issue-{E}ngineered {A}rticular {C}artilage {P}roperties}.
\newblock {\em Tissue Eng Part B Rev}, 24(5):345--358, 10 2018.

\bibitem{Caddeo2017Review}
Silvia Caddeo, Monica Boffito, and Susanna Sartori.
\newblock Tissue engineering approaches in the design of healthy and
  pathological in vitro tissue models.
\newblock {\em Frontiers in Bioengineering and Biotechnology}, 5:40, 2017.

\bibitem{Schiavi2018}
J.~Schiavi, L.~Reppel, N.~Charif, N.~de~Isla, D.~Mainard, N.~Benkirane-Jessel,
  J.~F. Stoltz, R.~Rahouadj, and C.~Huselstein.
\newblock {{M}echanical stimulations on human bone marrow mesenchymal stem
  cells enhance cells differentiation in a three-dimensional layered scaffold}.
\newblock {\em J Tissue Eng Regen Med}, 12(2):360--369, 02 2018.

\bibitem{Panadero2016}
J.~A. Panadero, S.~Lanceros-Mendez, and J.~L. Ribelles.
\newblock {{D}ifferentiation of mesenchymal stem cells for cartilage tissue
  engineering: {I}ndividual and synergetic effects of three-dimensional
  environment and mechanical loading}.
\newblock {\em Acta Biomater}, 33:1--12, Mar 2016.

\bibitem{Weaver2010Review}
C.~Frantz, K.~M. Stewart, and V.~M. Weaver.
\newblock {{T}he extracellular matrix at a glance}.
\newblock {\em J Cell Sci}, 123(Pt 24):4195--4200, Dec 2010.

\bibitem{Capulli2016Review}
A.~K. Capulli, L.~A. MacQueen, S.~P. Sheehy, and K.~K. Parker.
\newblock {{F}ibrous scaffolds for building hearts and heart parts}.
\newblock {\em Adv Drug Deliv Rev}, 96:83--102, Jan 2016.

\bibitem{Stocco2018}
Thiago~D. Stocco, Nicole~J. Bassous, Siqi Zhao, Alessandro E.~C. Granato,
  Thomas~J. Webster, and Anderson~O. Lobo.
\newblock Nanofibrous scaffolds for biomedical applications.
\newblock {\em Nanoscale}, 10:12228--12255, 2018.

\bibitem{yu2016decellularized}
Yaling Yu, Ali Alkhawaji, Yuqiang Ding, and Jin Mei.
\newblock Decellularized scaffolds in regenerative medicine.
\newblock {\em Oncotarget}, 7(36):58671, 2016.

\bibitem{sill2008electrospinning}
Travis~J Sill and Horst~A Von~Recum.
\newblock Electrospinning: applications in drug delivery and tissue
  engineering.
\newblock {\em Biomaterials}, 29(13):1989--2006, 2008.

\bibitem{agarwal2009progress}
Seema Agarwal, Joachim~H Wendorff, and Andreas Greiner.
\newblock Progress in the field of electrospinning for tissue engineering
  applications.
\newblock {\em Advanced Materials}, 21(32-33):3343--3351, 2009.

\bibitem{wang20203d}
Chong Wang, Wei Huang, Yu~Zhou, Libing He, Zhi He, Ziling Chen, Xiao He, Shuo
  Tian, Jiaming Liao, Bingheng Lu, et~al.
\newblock 3d printing of bone tissue engineering scaffolds.
\newblock {\em Bioactive materials}, 5(1):82--91, 2020.

\bibitem{Zhao2020}
Z.~Zhao, C.~Fan, F.~Chen, Y.~Sun, Y.~Xia, A.~Ji, and D.~A. Wang.
\newblock {{P}rogress in {A}rticular {C}artilage {T}issue {E}ngineering: {A}
  {R}eview on {T}herapeutic {C}ells and {M}acromolecular {S}caffolds}.
\newblock {\em Macromol Biosci}, 20(2):e1900278, 02 2020.

\bibitem{nair2021retina}
Deepthi S~Rajendran Nair, Magdalene~J Seiler, Kahini~H Patel, Vinoy Thomas,
  Juan Carlos~Martinez Camarillo, Mark~S Humayun, and Biju~B Thomas.
\newblock Tissue engineering strategies for retina regeneration.
\newblock {\em Applied Sciences}, 11(5):2154, 2021.

\bibitem{choi2019ivd}
Youngjoo Choi, Min~Hee Park, and Kangwon Lee.
\newblock Tissue engineering strategies for intervertebral disc treatment using
  functional polymers.
\newblock {\em Polymers}, 11(5):872, 2019.

\bibitem{bhardwaj20183d_skin}
Nandana Bhardwaj, Dimple Chouhan, and Biman~B Mandal.
\newblock 3d functional scaffolds for skin tissue engineering.
\newblock In {\em Functional 3D tissue engineering scaffolds}, pages 345--365.
  Elsevier, 2018.

\bibitem{vasiliadis2020tendon}
Angelo~V Vasiliadis and Konstantinos Katakalos.
\newblock The role of scaffolds in tendon tissue engineering.
\newblock {\em Journal of functional biomaterials}, 11(4):78, 2020.

\bibitem{devillard2021vascular}
Chlo{\'e}~D Devillard and Christophe~A Marquette.
\newblock Vascular tissue engineering: Challenges and requirements for an ideal
  large scale blood vessel.
\newblock {\em Frontiers in Bioengineering and Biotechnology}, page 913, 2021.

\bibitem{ghassemi2018bone}
Toktam Ghassemi, Azadeh Shahroodi, Mohammad~H Ebrahimzadeh, Alireza Mousavian,
  Jebraeel Movaffagh, and Ali Moradi.
\newblock Current concepts in scaffolding for bone tissue engineering.
\newblock {\em Archives of bone and joint surgery}, 6(2):90, 2018.

\bibitem{liebschner2005mechanical}
Michael Liebschner, Brandon Bucklen, and Matthew Wettergreen.
\newblock Mechanical aspects of tissue engineering.
\newblock In {\em Seminars in Plastic Surgery}, volume~19, pages 217--228.
  Copyright{\copyright} 2005 by Thieme Medical Publishers, Inc., 333 Seventh
  Avenue, New~…, 2005.

\bibitem{buffinton2015uniaxialElongation}
Christine~Miller Buffinton, Kelly~J Tong, Roberta~A Blaho, Elise~M Buffinton,
  and Donna~M Ebenstein.
\newblock Comparison of mechanical testing methods for biomaterials: Pipette
  aspiration, nanoindentation, and macroscale testing.
\newblock {\em journal of the mechanical behavior of biomedical materials},
  51:367--379, 2015.

\bibitem{sacks2000biaxial}
Michael~S Sacks.
\newblock Biaxial mechanical evaluation of planar biological materials.
\newblock {\em Journal of elasticity and the physical science of solids},
  61(1):199--246, 2000.

\bibitem{ghezelbash2021shear}
Farshid Ghezelbash, Amir~Hossein Eskandari, Aboulfazl Shirazi-Adl, Morteza
  Kazempour, Javad Tavakoli, Mostafa Baghani, and John~J Costi.
\newblock Modeling of human intervertebral disc annulus fibrosus with complex
  multi-fiber networks.
\newblock {\em Acta Biomaterialia}, 123:208--221, 2021.

\bibitem{ogden1997non}
R.W. Ogden.
\newblock {\em Non-linear Elastic Deformations}.
\newblock Dover Civil and Mechanical Engineering. Dover Publications, 1997.

\bibitem{HolzapfelGasserOgden2000}
G.A. Holzapfel, T.C. Gasser, and R.W. Ogden.
\newblock A new constitutive framework for arterial wall mechanics and a
  comparative study of material models.
\newblock {\em Journal of Elasticity}, 61:1--48, 2000.

\bibitem{FedericoGasser2010}
S.~Federico and T.~C. Gasser.
\newblock {{N}onlinear elasticity of biological tissues with statistical fibre
  orientation}.
\newblock {\em J R Soc Interface}, 7(47):955--966, Jun 2010.

\bibitem{Shenoy2014b}
H.~Wang, A.~S. Abhilash, C.~S. Chen, R.~G. Wells, and V.~B. Shenoy.
\newblock {{L}ong-range force transmission in fibrous matrices enabled by
  tension-driven alignment of fibers}.
\newblock {\em Biophys J}, 107(11):2592--2603, Dec 2014.

\bibitem{Jones2015PNAS}
C.~A. Jones, M.~Cibula, J.~Feng, E.~A. Krnacik, D.~H. McIntyre, H.~Levine, and
  B.~Sun.
\newblock {{M}icromechanics of cellularized biopolymer networks}.
\newblock {\em Proc Natl Acad Sci U S A}, 112(37):E5117--5122, Sep 2015.

\bibitem{Rizvi2012ActaBiomat}
M.S. Rizvi, P.~Kumar, D.S. Katti, and A.~Pal.
\newblock Mathematical model of mechanical behavior of micro/nanofibrous
  materials designed for extracellular matrix substitutes.
\newblock {\em Acta Biomaterialia}, 8(11):4111--4122, 2012.

\bibitem{Rizvi2014JMBBM}
Mohd~Suhail Rizvi and Anupam Pal.
\newblock Statistical model for the mechanical behavior of the tissue
  engineering non-woven fibrous matrices under large deformation.
\newblock {\em Journal of the Mechanical Behavior of Biomedical Materials},
  37:235--250, 2014.

\bibitem{Rizvi2016BMM}
M.~S. Rizvi, A.~Pal, and S.~L. Das.
\newblock {{S}tructure-induced nonlinear viscoelasticity of non-woven fibrous
  matrices}.
\newblock {\em Biomech Model Mechanobiol}, 15(6):1641--1654, 12 2016.

\bibitem{Razavi2020}
P.~Chavoshnejad and M.~J. Razavi.
\newblock {{E}ffect of the {I}nterfiber {B}onding on the {M}echanical
  {B}ehavior of {E}lectrospun {F}ibrous {M}ats}.
\newblock {\em Sci Rep}, 10(1):7709, May 2020.

\bibitem{Eichinger2021}
J.~F. Eichinger, M.~J. Grill, I.~D. Kermani, R.~C. Aydin, W.~A. Wall, J.~D.
  Humphrey, and C.~J. Cyron.
\newblock {{A} computational framework for modeling cell-matrix interactions in
  soft biological tissues}.
\newblock {\em Biomech Model Mechanobiol}, 20(5):1851--1870, Oct 2021.

\bibitem{Tyznik2019}
Stephen Tyznik and Jacob Notbohm.
\newblock Length scale dependent elasticity in random three-dimensional fiber
  networks.
\newblock {\em Mechanics of Materials}, 138:103155, 2019.

\bibitem{MacKintosh2015}
A.~J. Licup, S.~Münster, A.~Sharma, M.~Sheinman, L.~M. Jawerth, B.~Fabry,
  D.~A. Weitz, and F.~C. MacKintosh.
\newblock {{S}tress controls the mechanics of collagen networks}.
\newblock {\em Proc Natl Acad Sci U S A}, 112(31):9573--9578, Aug 2015.

\bibitem{Rohanifar2020}
H.~Hatami-Marbini and M.~Rohanifar.
\newblock {{M}echanical properties of subisostatic random networks composed of
  nonlinear fibers}.
\newblock {\em Soft Matter}, 16(30):7156--7164, Aug 2020.

\bibitem{JanmeyShenoy2019}
E.~Ban, H.~Wang, J.~M. Franklin, J.~T. Liphardt, P.~A. Janmey, and V.~B.
  Shenoy.
\newblock {{S}trong triaxial coupling and anomalous {P}oisson effect in
  collagen networks}.
\newblock {\em Proc Natl Acad Sci U S A}, 116(14):6790--6799, 04 2019.

\bibitem{raghavan2011control}
Bharath~K Raghavan and Douglas~W Coffin.
\newblock Control of inter-fiber fusing for nanofiber webs via electrospinning.
\newblock {\em Journal of Engineered Fibers and Fabrics},
  6(4):155892501100600401, 2011.

\bibitem{Grimmer2018}
P.~Grimmer and J.~Notbohm.
\newblock {{D}isplacement {P}ropagation in {F}ibrous {N}etworks {D}ue to
  {L}ocal {C}ontraction}.
\newblock {\em J Biomech Eng}, 140(4), 04 2018.

\bibitem{Proestaki2019}
M.~Proestaki, A.~Ogren, B.~Burkel, and J.~Notbohm.
\newblock {Modulus of Fibrous Collagen at the Length Scale of a Cell}.
\newblock {\em Exp Mech}, 59:1323--1334, 2019.

\bibitem{Goren2020BiophysJ}
S.~Goren, Y.~Koren, X.~Xu, and A.~Lesman.
\newblock {{E}lastic {A}nisotropy {G}overns the {R}ange of {C}ell-{I}nduced
  {D}isplacements}.
\newblock {\em Biophys J}, 118(5):1152--1164, 03 2020.

\bibitem{Mann2019}
A.~Mann, R.~S. Sopher, S.~Goren, O.~Shelah, O.~Tchaicheeyan, and A.~Lesman.
\newblock {{F}orce chains in cell-cell mechanical communication}.
\newblock {\em J R Soc Interface}, 16(159):20190348, 10 2019.

\bibitem{Burkel2018PRE}
B.~Burkel, M.~Proestaki, S.~Tyznik, and J.~Notbohm.
\newblock {{H}eterogeneity and nonaffinity of cell-induced matrix
  displacements}.
\newblock {\em Phys Rev E}, 98(5), Nov 2018.

\bibitem{Hall2016PNAS}
M.~S. Hall, F.~Alisafaei, E.~Ban, X.~Feng, C.~Y. Hui, V.~B. Shenoy, and M.~Wu.
\newblock {{F}ibrous nonlinear elasticity enables positive mechanical feedback
  between cells and {E}{C}{M}s}.
\newblock {\em Proc Natl Acad Sci U S A}, 113(49):14043--14048, 12 2016.

\bibitem{Broedersz2018}
Y.~L. Han, P.~Ronceray, G.~Xu, A.~Malandrino, R.~D. Kamm, M.~Lenz, C.~P.
  Broedersz, and M.~Guo.
\newblock {{C}ell contraction induces long-ranged stress stiffening in the
  extracellular matrix}.
\newblock {\em Proc Natl Acad Sci U S A}, 115(16):4075--4080, 04 2018.

\bibitem{Shenoy2014}
A.~S. Abhilash, B.~M. Baker, B.~Trappmann, C.~S. Chen, and V.~B. Shenoy.
\newblock {{R}emodeling of fibrous extracellular matrices by contractile cells:
  predictions from discrete fiber network simulations}.
\newblock {\em Biophys J}, 107(8):1829--1840, Oct 2014.

\bibitem{DeSafran2008}
R.~De, A.~Zemel, and S.~A. Safran.
\newblock {{D}o cells sense stress or strain? {M}easurement of cellular
  orientation can provide a clue}.
\newblock {\em Biophys J}, 94(5):29--31, Mar 2008.

\bibitem{Panzetta2019PNAS}
Valeria Panzetta, Sabato Fusco, and Paolo~A. Netti.
\newblock Cell mechanosensing is regulated by substrate strain energy rather
  than stiffness.
\newblock {\em Proceedings of the National Academy of Sciences},
  116(44):22004--22013, 2019.

\bibitem{Electrospinning2019Review}
J.~Xue, T.~Wu, Y.~Dai, and Y.~Xia.
\newblock {Electrospinning and Electrospun Nanofibers: Methods, Materials, and
  Applications}.
\newblock {\em Chem Rev}, 119(8):5298--5415, 03 2019.

\bibitem{Luraghi2021Review}
Andrea Luraghi, Francesco Peri, and Lorenzo Moroni.
\newblock Electrospinning for drug delivery applications: A review.
\newblock {\em Journal of Controlled Release}, 334:463--484, 2021.

\bibitem{ryu2020thickness}
Hyun~Il Ryu, Min~Seok Koo, Seokjun Kim, Songkil Kim, Young-Ah Park, and
  Sang~Min Park.
\newblock Uniform-thickness electrospun nanofiber mat production system based
  on real-time thickness measurement.
\newblock {\em Scientific reports}, 10(1):1--10, 2020.

\bibitem{gurtin1982introduction}
Morton~E Gurtin.
\newblock {\em An introduction to continuum mechanics}.
\newblock Academic press, 1982.

\bibitem{courtney2006design}
Todd Courtney, Michael~S Sacks, John Stankus, Jianjun Guan, and William~R
  Wagner.
\newblock Design and analysis of tissue engineering scaffolds that mimic soft
  tissue mechanical anisotropy.
\newblock {\em Biomaterials}, 27(19):3631--3638, 2006.

\bibitem{storm2005}
Cornelis Storm, Jennifer~J Pastore, Fred~C MacKintosh, Tom~C Lubensky, and
  Paul~A Janmey.
\newblock Nonlinear elasticity in biological gels.
\newblock {\em Nature}, 435(7039):191--194, 2005.

\bibitem{Gasser2006}
T.~C. Gasser, R.~W. Ogden, and G.~A. Holzapfel.
\newblock {{H}yperelastic modelling of arterial layers with distributed
  collagen fibre orientations}.
\newblock {\em J R Soc Interface}, 3(6):15--35, Feb 2006.

\bibitem{gasser2006hyperelastic}
T~Christian Gasser, Ray~W Ogden, and Gerhard~A Holzapfel.
\newblock Hyperelastic modelling of arterial layers with distributed collagen
  fibre orientations.
\newblock {\em Journal of the royal society interface}, 3(6):15--35, 2006.

\bibitem{chen2007role}
Ming Chen, Prabir~K Patra, Steven~B Warner, and Sankha Bhowmick.
\newblock Role of fiber diameter in adhesion and proliferation of nih 3t3
  fibroblast on electrospun polycaprolactone scaffolds.
\newblock {\em Tissue engineering}, 13(3):579--587, 2007.

\bibitem{prasadh2018unraveling}
Somasundaram Prasadh and Raymond Chung~Wen Wong.
\newblock Unraveling the mechanical strength of biomaterials used as a bone
  scaffold in oral and maxillofacial defects.
\newblock {\em Oral Science International}, 15(2):48--55, 2018.

\bibitem{griffin2016biomechanical}
Michelle Griffin, Yaami Premakumar, Alexander Seifalian, Peter~Edward Butler,
  and Matthew Szarko.
\newblock Biomechanical characterization of human soft tissues using
  indentation and tensile testing.
\newblock {\em Journal of visualized experiments: JoVE}, (118), 2016.

\bibitem{Martino2018Review}
Fabiana Martino, Ana~R. Perestrelo, Vladimír Vinarský, Stefania Pagliari, and
  Giancarlo Forte.
\newblock Cellular mechanotransduction: From tension to function.
\newblock {\em Frontiers in Physiology}, 9:824, 2018.

\bibitem{PodioGuidugli2014}
P.~Podio-Guidugli and A.~Favata.
\newblock {\em The Kelvin Problem}, pages 115--129.
\newblock Springer International Publishing, Cham, 2014.

\bibitem{beroz2017physical}
Farzan Beroz, Louise~M Jawerth, Stefan M{\"u}nster, David~A Weitz, Chase~P
  Broedersz, and Ned~S Wingreen.
\newblock Physical limits to biomechanical sensing in disordered fibre
  networks.
\newblock {\em Nature communications}, 8(1):1--11, 2017.

\bibitem{gjorevski2012}
Nikolce Gjorevski and Celeste~M Nelson.
\newblock Mapping of mechanical strains and stresses around quiescent
  engineered three-dimensional epithelial tissues.
\newblock {\em Biophysical journal}, 103(1):152--162, 2012.

\bibitem{Janmey2020Review}
P.~A. Janmey, D.~A. Fletcher, and C.~A. Reinhart-King.
\newblock {{S}tiffness {S}ensing by {C}ells}.
\newblock {\em Physiol Rev}, 100(2):695--724, 04 2020.

\bibitem{DeSantis2011}
G.~De~Santis, A.~B. Lennon, F.~Boschetti, B.~Verhegghe, P.~Verdonck, and P.~J.
  Prendergast.
\newblock {{H}ow can cells sense the elasticity of a substrate? {A}n analysis
  using a cell tensegrity model}.
\newblock {\em Eur Cell Mater}, 22:202--213, Oct 2011.

\bibitem{rosenbluth2006force}
Michael~J Rosenbluth, Wilbur~A Lam, and Daniel~A Fletcher.
\newblock Force microscopy of nonadherent cells: a comparison of leukemia cell
  deformability.
\newblock {\em Biophysical journal}, 90(8):2994--3003, 2006.

\bibitem{mathur2001endothelial}
Anshu~B Mathur, Amy~M Collinsworth, William~M Reichert, William~E Kraus, and
  George~A Truskey.
\newblock Endothelial, cardiac muscle and skeletal muscle exhibit different
  viscous and elastic properties as determined by atomic force microscopy.
\newblock {\em Journal of biomechanics}, 34(12):1545--1553, 2001.

\bibitem{collet2005elasticity}
Jean-Philippe Collet, Henry Shuman, Robert~E Ledger, Seungtaek Lee, and John~W
  Weisel.
\newblock The elasticity of an individual fibrin fiber in a clot.
\newblock {\em Proceedings of the National Academy of Sciences},
  102(26):9133--9137, 2005.

\bibitem{oral2011measuring}
Imran Oral, Hatice Guzel, and Gulnare Ahmetli.
\newblock Measuring the young’s modulus of polystyrene-based composites by
  tensile test and pulse-echo method.
\newblock {\em Polymer bulletin}, 67(9):1893--1906, 2011.

\bibitem{zhu2008materials}
TT~Zhu, AJ~Bushby, and DJ~Dunstan.
\newblock Materials mechanical size effects: a review.
\newblock {\em Materials Technology}, 23(4):193--209, 2008.

\bibitem{iturri2017afm}
Jagoba Iturri and Jos{\'e}~L Toca-Herrera.
\newblock Characterization of cell scaffolds by atomic force microscopy.
\newblock {\em Polymers}, 9(8):383, 2017.

\bibitem{discher2005tissue}
Dennis~E Discher, Paul Janmey, and Yu-li Wang.
\newblock Tissue cells feel and respond to the stiffness of their substrate.
\newblock {\em Science}, 310(5751):1139--1143, 2005.

\bibitem{pelham1999high}
Robert~J Pelham~Jr and Yu-li Wang.
\newblock High resolution detection of mechanical forces exerted by locomoting
  fibroblasts on the substrate.
\newblock {\em Molecular biology of the cell}, 10(4):935--945, 1999.

\bibitem{burton1999keratocytes}
Kevin Burton, Jung~H Park, and D~Lansing Taylor.
\newblock Keratocytes generate traction forces in two phases.
\newblock {\em Molecular biology of the cell}, 10(11):3745--3769, 1999.

\bibitem{yin2005myocite}
Shizhuo Yin, Xueqian Zhang, Chun Zhan, Juntao Wu, Jinchao Xu, and Joseph
  Cheung.
\newblock Measuring single cardiac myocyte contractile force via moving a
  magnetic bead.
\newblock {\em Biophysical journal}, 88(2):1489--1495, 2005.

\bibitem{Janmey2009Review}
S.~Y. Tee, A.~R. Bausch, and P.~A. Janmey.
\newblock {{T}he mechanical cell}.
\newblock {\em Curr Biol}, 19(17):R745--748, Sep 2009.

\bibitem{gupta2016single}
Mukund Gupta, Bryant Doss, Chwee~Teck Lim, Raphael Voituriez, and Benoit
  Ladoux.
\newblock Single cell rigidity sensing: a complex relationship between focal
  adhesion dynamics and large-scale actin cytoskeleton remodeling.
\newblock {\em Cell adhesion \& migration}, 10(5):554--567, 2016.

\bibitem{barbier2009friction}
Carine Barbier, R{\'e}my Dendievel, and David Rodney.
\newblock Role of friction in the mechanics of nonbonded fibrous materials.
\newblock {\em Physical Review E}, 80(1):016115, 2009.

\bibitem{bauchau2009euler}
Oliver~A Bauchau and James~I Craig.
\newblock Euler-bernoulli beam theory.
\newblock In {\em Structural analysis}, pages 173--221. Springer, 2009.

\bibitem{ochsnerclassical}
Andreas {\"O}chsner.
\newblock Classical beam theories of structural mechanics.

\end{thebibliography}

\end{document}